\documentclass[12pt,subeqn,a4paper]{article}

\usepackage{amssymb,amsmath,amsfonts,amsthm, amscd, mathrsfs,helvet,multirow}
\usepackage{bm}
\usepackage{graphicx,verbatim}
\usepackage{psfrag}
\usepackage[all]{xy}

\usepackage{appendix}
\numberwithin{equation}{section}
\numberwithin{figure}{section}

\newcommand{\F}{\mathcal{F}}
\DeclareMathOperator{\im}{Im}
\DeclareMathOperator{\re}{Re}
\DeclareMathOperator{\tr}{tr}
\newcommand{\inv}{\left(\im \tau^{-1}\right)}

\def\r2{\sqrt{2}}

\begin{document}
%zzz
\newcommand{\nd}[1]{/\hspace{-0.5em} #1}
\begin{titlepage}
\begin{flushright}
{\bf December 2014} \\ 
DAMTP-2014-93\\
%SWAT-\\ 
%hep-th/yymmnnn \\
\end{flushright}
\begin{centering}
\vspace{.2in}
 {\large {\bf Instantons, Integrability and Discrete 
     Light-Cone Quantisation}}

\vspace{.3in}

Nick Dorey and Andrew Singleton\\
\vspace{.1 in}
DAMTP, Centre for Mathematical Sciences \\ 
University of Cambridge, Wilberforce Road \\ 
Cambridge CB3 0WA, UK \\
\vspace{.2in}
%and \\ 
%\vspace{.2in}
%
%
\vspace{.4in}
{\bf Abstract} \\

\end{centering}
We study supersymmetric quantum mechanics on the moduli space of
Yang-Mills instantons on $\mathbb{R}^{2}\times T^{2}$ and its
application to the discrete light-cone quantisation (DLCQ) of ${\cal N}=4$
SUSY Yang-Mills. In the presence
of a target space magnetic field, the model has a discrete spectrum with
the wavefunctions of generic energy eigenstates supported away from
the singular points of the moduli space. 
The corresponding Hamiltonian is part of an ${\bf osp}(1,1|4)$
superalgebra which enlarges to ${\bf su}(1,1|4)$ superconformal invariance in
the sector corresponding to the ${\cal
  N}=4$ theory. The Hamiltonian is
isospectral to the light-cone dilatation operator of the ${\cal N}=4$
theory in this sector. 
The model also has an interesting scaling limit where
it becomes integrable. We determine the semiclassical spectrum in this
limit. We discuss a possible approach to constructing the 
dilatation operator of ${\cal N}=4$ supersymmetric Yang-Mills theory
in DLCQ.

%\vspace{.05in}
%\baselineskip=.3in
\end{titlepage}
\paragraph{}
\section{Introduction}
\paragraph{}
The developments of the last few years strongly suggest the existence
of hidden symmetries in non-abelian gauge theory which are not
manifest in the standard Lagrangian formulation of the theory. 
This is particularly apparent in the case of ${\cal N}=4$
supersymmetric Yang-Mills theory where remarkable simplifications
occur in perturbative computations of 
operator dimensions, scattering amplitudes and other observables. 
In the planar limit, there is overwhelming evidence
that the theory is exactly integrable and the assumption of integrability
leads to an exact solution for the anomalous dimensions of all 
local operators in this limit \cite{BigReview:2010}. Despite this, the origin of
integrability and of the corresponding symmetries in gauge theory remains
obscure. In particular, the spin chain Hamiltonian is corrected by longer
range interactions order by order in perturbation theory 
and its persistent integrability at each new order seems miraculous. 
The existing bootstrap solution of the spectral problem provides
detailed information about the eigenvalues of the planar dilatation
operator but not the underlying Hamiltonian or the corresponding wave
functions.  
\paragraph{}
For the reasons described above, it is natural to look for a
formulation of the ${\cal N}=4$ theory in which some or all of the hidden
symmetries become manifest\footnote{One possible candidate is provided 
by the worldsheet theory of the dual string which is
classically integrable. However, the first-principles 
quantisation of the worldsheet $\sigma$-model
for the Green-Schwarz string in $AdS_{5}\times S^{5}$ is a hard
problem which remains unsolved. The main obstacle is in finding a
discretised version of the worldsheet theory in which integrability is
preserved. For recent progress in this direction, see however \cite{Delduc:Magro:Vicedo:2012}.}. 
With this goal in mind, we will study the ${\cal N}=4$ theory with
gauge group $SU(N)$ in Discrete
Light-Cone Quantisation (DLCQ) following the proposal of 
\cite{Ganor:Sethi:1997, Kapustin:Sethi:1998}. In this formalism, one
studies the theory compactified on a light-like circle of fixed radius
$R_{-}$, so that the corresponding null momentum is quantised in
integer units: $p_{+}=K/R_{-}$, $K\in \mathbb{N}$. As usual, restricting 
the theory to a sector with $K$ units of null momentum reduces a field
theory to a finite dimensional quantum mechanical model with a number
of degrees of freedom which grows linearly with $K$. An obvious hope
is that the finite-dimensional model might itself be integrable or,
more realistically, might become so for large $N$. 
In this paper, we will give a precise formulation of the
relevant quantum mechanical model and show that it 
is indeed at least semiclassically integrable in a
certain limit (even for finite $N$). The relation to the
integrability of the planar ${\cal N}=4$ theory remains to be
understood. Along the way we will also perform some basic checks on
the proposed DLCQ description.
 \paragraph{}        
As we review below, the DLCQ decription of the ${\cal N}=4$ theory 
with gauge group $SU(N)$ 
in the sector with $K$ units of null momentum is given by supersymmetric
quantum mechanics on the moduli space of $K$ instantons of an
auxiliary $SU(N)$ gauge theory formulated on $\mathbb{R}^{2}\times
T^{2}$. More precisely, we should take a limit where the area of the
torus goes to zero. The model also has additional parameters which
correspond to light-like Wilson lines breaking the
gauge group of the DLCQ theory down to its Cartan subgroup. 
This proposal of \cite{Ganor:Sethi:1997, Kapustin:Sethi:1998} has
its origin in the realisation of ${\cal N}=4$ SUSY Yang-Mills as an IR
fixed point arising when the six-dimensional $(2,0)$ theory is compactified to
four dimensions on a torus \cite{Witten:1995}. 
An attractive feature of this approach is
that dependence on the gauge theory coupling is encoded geometrically 
in the complex structure of the torus. 
\paragraph{}
In this paper we first establish some basic results about the proposed DLCQ
description. The moduli space of $K$ $SU(N)$ 
instantons on $\mathbb{R}^{2}\times T^{2}$
is a hyper-K\"{a}hler manifold\footnote{The manifold has singularities
  which we will discuss below.} of real dimension $4r=4(KN-N+1)$ which
we will denote as $\mathcal{M}$. 
Although the full hyper-K\"{a}hler 
metric on $\mathcal{M}$ is complicated, it approaches a 
simple analytic form known as the 
{\em semi-flat metric} in the limit relevant for describing the 
${\cal N}=4$ theory. Building
on the results of our previous paper \cite{ND:AS:2014}, we construct the
relevant quantum mechanical $\sigma$-model with target space $\mathcal{M}$
equipped with the semi-flat metric. An important qualitative feature
of the moduli space is that each $SU(N)$ instanton 
splits into $N$ constituents or ``partons'' moving on
$\mathbb{R}^{2}\times T^{2}$. The model has a weak coupling limit
where the semi-flat metric becomes flat and the system consists of
$KN$ free partons\footnote{More precisely there are $K$ identical
  partons in each of $N$ species with $N-1$ linear constraints on the relative
  positions of the center of mass of 
  each species.}. Away from this limit, the partons
interact with each other through the curvature of the 
moduli space metric. Despite this, the semi-flat $\sigma$-model 
has $2r$ conserved charges\footnote{As we explain below, $\mathcal{M}$
  can be understood as a torus fibration over a special K\"{a}hler
  base. Special K\"{a}hler geometry provides a symplectic pairing
  between ``electric'' and ``magnetic'' cycles of the fibre denoted by
  the labels $e$ and $m$ respectively.} $\{Q_{e\,I},Q_{m}^{I}\}$,
$I=1,2,\ldots,r$, corresponding to the momenta of {\em individual} 
partons in the two compact directions. Although this is only half the
number of conserved charges needed to make the $\sigma$-model
integrable, we will describe an interesting limit where the system
nevertheless becomes integrable. 
\paragraph{}
In the context of DLCQ, $Q_{e\, I}$ and
$Q^{I}_{m}$ correspond to momenta carried by the tower of Kaluza-Klein
states in the compactification of the $(2,0)$-theory on
$\mathbb{R}^{3,1}\times T^{2}$. In its DLCQ description, the $(2,0)$
theory has an $SU(N)$ gauge symmetry and generic states with momenta in the
compact directions also carry the corresponding gauge charges. 
The gauge-invariant states of
the ${\cal N}=4$ theory should correspond to the sector of the
$\sigma$-model with  $Q_{e\, I}=Q^{I}_{m}=0$. We present two pieces of
evidence in favour of this identification. First, we show that the
supersymmetry of the $\sigma$-model is enlarged to an $SU(1,1|4)$ 
superconformal invariance in this sector. 
This precisely matches the unbroken
superconformal invariance of ${\cal N}=4$ supersymmetric
Yang-Mills theory compactified on a light-like circle. In particular,
the dilatation operator of the quantum mechanical model is related to
the ``light-cone'' dilatation operator of the ${\cal N}=4$ theory.  
Second, in a weak-coupling limit where the curvature of the
target space becomes small, the partons correspond to
perturbative quanta of an $SU(N)$ gauge theory. In the sector with      
$Q_{e\, I}=Q^{I}_{m}=0$, the spectrum precisely coincides with that of
the free ${\cal N}=4$ gauge theory on a light-like circle.  
\paragraph{}
The results described above are provisional in character 
because the instanton moduli-space is singular. 
In addition to the familiar small instanton singularities, more severe
singularities arise in the semi-flat limit relevant for describing the
${\cal N}=4$ theory. In order to obtain a useful description of ${\cal
  N}=4$ SUSY Yang-Mills we need a prescription for dealing with these
singularities. To this end we consider a deformation of the
$\sigma$-model corresponding to the introduction of a target space
magnetic field. As we discuss in Section 5, the 
new model corresponds to a DLCQ description of the
$(2,0)$-theory compactified on a certain $T^{2}$ bundle over a
four-dimensional pp-wave geometry. 
The deformation has no effect on states in the sector
with zero parton momenta which should contain the states of the ${\cal
  N}=4$ theory. In particular it leaves the $SU(1,1|4)$ superconformal
symmetry of that sector unbroken. However, we will show that 
it has several interesting effects on generic states of the full model:
\paragraph{}
{\bf 1:} In the presence of the deformation, the full $\sigma$-model
has a non-trivial superconformal invariance. In particular the
new Hamiltonian 
is part of an ${\bf osp}(1,1|4)$ superconformal   
algebra. This is a subalgebra of the ${\bf su}(1,1|4)$ invariance of
the sector of zero parton momenta. 
\paragraph{}
{\bf 2:} The deformed theory is characterised by a new dimensionless
parameter $\rho$ measuring the strength of the applied
magnetic field. In the sector with $Q_{e\, I}=Q_{m}^{I}=0$, changing $\rho$
simply corresponds to a change of basis of the conformal algebra which leaves
the spectrum invariant. For $Q_{e\, I}, Q^{I}_{m}\neq 0$, the spectrum
depends non-trivially on $\rho$. 
\paragraph{}
{\bf 3:} For $\rho>>1$, 
wavefunctions of generic states are supported away from the
singular points of $\mathcal{M}$. The resulting spectrum is discrete.  
\paragraph{}
{\bf 4:} In the limit $\rho\rightarrow \infty$, for $Q_{e\, I}, Q^{I}_{m}\neq
0$, the model becomes (at least) semiclassically integrable. We determine the
semiclassical spectrum of the model in this limit. At weak coupling, the
spectrum can be related to that of an integrable spin chain.  
\paragraph{}
In the limit $\rho\rightarrow\infty$, it seems possible that the model
has full quantum integrability and that its spectrum can be
determined exactly. The most interesting
question, however, is what relation, if any, our results have to the
integrability of the planar ${\cal N}=4$ theory. In the final
section of the paper we make some preliminary remarks which should
serve as a starting point for further investigation.  
In particular we note that, 
{\em after} taking $\rho\rightarrow\infty$, the resulting
description in terms of an integrable system appears 
to extend smoothly to states with $Q_{e\, I},Q^{I}_{m}=0$ suggesting that it
may be possible to extract the operator dimensions of the ${\cal N}=4$
theory. This will be investigated elswhere \cite{ND:future}. 
\paragraph{}
The paper is organised as follows. In Section 2, we review the
properties of the relevant instanton moduli-space, $\mathcal{M}$. In
Section 3 we construct the quantum mechanical $\sigma$-model with
target space $\mathcal{M}$ and its deformation by an external magnetic
field. In Section 4, we describe the specific application to the DLCQ
of ${\cal N}=4$ supersymmetric Yang-Mills theory. In Section 5 we
propose a spacetime interpretation for the magnetic
deformation. Finally in Section 6, we discuss prospects for extracting
physical observables of the ${\cal N}=4$ theory from this approach.     
  
\section{The instanton moduli space}
\paragraph{}
Following the proposal of \cite{Ganor:Sethi:1997,
  Kapustin:Sethi:1998}, we will consider supersymmetric quantum mechanics on the
moduli space, $\mathcal{M}$,  
of $K$ instantons of an $SU(N)$ Yang-Mills theory living
on $\mathbb{R}^{2}\times T^{2}$. The torus has complex structure
parameter $\tau_{\rm cl}$ and area $\mathcal{A}=4\pi^{2}{\rm
  Im}\tau_{\rm cl}\ell^{2}$ where $\ell$ is a parameter with the
dimensions of length. 
We also introduce Wilson lines for the $SU(N)$ gauge field on 
$T^{2}$. The Wilson line is specified by choosing $N$ marked points on
the dual torus $\hat{T}^2$ (up to an overall translation). 
The instanton moduli space 
is a hyper-K\"{a}hler manifold and bosonic quantum mechanics on $\mathcal{M}$
therefore admits a fermionic completion with $\mathcal{N}=(4,4)$
supersymmetry.  
In the remainder of this section we review 
several known results about the moduli space. The reader
should consult the original references \cite{Kapustin:Sethi:1998, Nekrasov:1995, 
Seiberg:Witten:1996, 
Kapustin:1998, ND:Hollowood:Kumar:2001} for further details. 
\paragraph{}
In fact the  hyper-K\"{a}hler 
manifold ${\cal M}$ arises in two distinct ways 
as the vacuum moduli space of an auxiliary supersymmetric gauge
theory. First, the ADHM-Nahm construction of the moduli space 
can be interpreted as an infinite-dimensional hyper-K\"{a}hler
quotient. The same quotient construction describes the Higgs branch of
a $U(K)$ gauge theory with eight supercharges 
on $\hat{T}^{2}\times \mathbb{R}^{2,1}$ with 
impurities localised at the $N$ marked points on the torus. 
As the theory is three-dimensional
in the IR, we can also apply the mirror symmetry of 
\cite{Intriligator:Seiberg:1996} to
realise ${\cal M}$ as the Coulomb branch of another theory with
eight supercharges and 
three non-compact dimensions. We will call these two constructions
the ``Higgs branch'' and ``Coulomb branch'' descriptions of ${\cal
  M}$. We will now briefly review both these descriptions.  
\subsection{The Coulomb branch description}
\paragraph{}
The Coulomb branch description starts from an ${\cal N}=2$
quiver gauge theory in four dimensions. The 
corresponding quiver diagram coincides with the Dynkin diagram for the affine
Lie algebra $\hat{A}_{N-1}$. The model coincides with the ``elliptic
quiver'' introduced and solved in \cite{Witten:1997}. 
Thus we consider an 
${\cal N}=2$ supersymmetric gauge theory
with gauge group,     
\begin{eqnarray}
G & = & U(1)_{D}\times \prod_{j=1}^{N} \, SU(K)_{j},
\label{eqn:quivergroup} 
\end{eqnarray}     
In addition to an ${\cal N}=2$ vector multiplet for each $SU(K)$
factor in $G$, the theory contains hypermultiplets in the
bifundamental representation of adjacent factors. 
Thus we have a
hypermultiplet in the $(\bar{\bf K}, {\bf K})$ of  
$SU(K)_{j}\times SU(K)_{j+1}$ for $j=1,2,\ldots,N$ with the
identification $SU(K)_{N+1}\simeq SU(K)_{1}$. All matter 
fields are neutral under a diagonal factor $U(1)_{D}$ which decouples. 
The complexified gauge coupling of the
diagonal $U(K)=U(1)_{D}\times SU(K)_{D}/\mathbb{Z}_{K}$ 
coincides with the complex
structure parameter $\tau_{\rm cl}$ of the torus in the ambient spacetime of the 
$SU(N)$ instantons. The $N-1$ off-diagonal gauge couplings are encoded
in the relative positions of $N$ marked points on $\hat{T}^{2}$. 
The $\beta$-function of each gauge coupling vanishes
and the four-dimensional theory has $N$ exactly marginal 
couplings (including $\tau_{\rm cl}$).     
\paragraph{}
The theory has a Coulomb branch where the complex scalars in the
vector multiplet acquire non-zero vacuum expectation values and Higgs
the gauge group down to its Cartan subgroup $U(1)^{r}$ where
$r=KN-N+1$. The low-energy physics on the Coulomb branch is governed
by a complex curve $\Sigma$ of genus $r$ \cite{Witten:1997}. 
To describe the curve we start with a canonical complex torus, 
\begin{eqnarray}
E(\tau) & = & \frac{\mathbb{C}}{2\omega_{1}\mathbb{Z}\oplus 
2\omega_{2}\mathbb{Z}} 
\nonumber
\end{eqnarray}
where $\omega_{1}=i\pi$ and $\omega_{2}=i\pi \tau_{\rm cl}$. We also specify
$N$ marked points $z=Z_{j}$ for $j=1,2,\ldots,N$. 
The gauge coupling of the factor $SU(N)_{j}$ in $G$ is identified as 
$\tau_{j}=(Z_{j+1}-Z_{j})/2\omega_{1}$ with the identification 
$Z_{N+1}=Z_{1}$. 
\paragraph{}  
The curve $\Sigma$ is a $K$-fold branched cover of $E(\tau)$ 
which is embedded in $\mathbb{C}\times E(\tau)$ as follows. Choosing
coordinates $v\in \mathbb{C}$ and $z\in E(\tau)$, $\Sigma$ is
specified by a polynomial equation of the form, 
\begin{eqnarray}
F(v,z) & =  & v^{K}-f_{1}(z)v^{K-1}+
f_{2}(z)v^{K-2}-\ldots+(-1)^{K}f_{K}(z)=0.
\label{curve}
\end{eqnarray}
For each $z\in E(\tau)$, the equation for $v$ has $K$ roots
corresponding to the $K$ branches of the cover. The $K$ coefficient
functions are constrained as follows: $f_{1}(z)$ is constant
while $f_{a}(z)$ for $a>1$ are meromorphic on $E(\tau)$ with simple
poles at the points $z=Z_{i}$, for $i=1,2,\ldots,N$ 
and no other singularities. We can solve
these conditions in terms of elliptic functions\footnote{In the following 
$\sigma(z)$, $\xi(z)=\sigma'(z)/\sigma(z)$ and
$\mathcal{P}(z)=-\xi'(z)$ denote the standard Weierstrass
functions} as follows, 
\begin{eqnarray}
f_{1}(z) & \equiv & {\cal H}^{(0)}_{1} \nonumber \\ 
f_{a}(z) & =& \sum_{i=1}^{N}\, {\cal
  H}^{(i)}_{a}\xi\left(z-Z_{i}\right)\,+\, {\cal H}_{a}^{(0)}\qquad{}
{\rm for} \,\,\,\,a=2,3,\ldots,K. \label{param}  
\end{eqnarray} 
The $(K-1)(N+1)+1$ complex
coefficients $\{{\cal H}^{(i)}_{a}$,  ${\cal H}^{(0)}_{a}\}$ 
are subject to the constraints, 
\begin{eqnarray}
 \sum_{i=1}^{N}\, {\cal
  H}^{(i)}_{a} & = & 0 \qquad{}
{\rm for} \,\,\,\,a=2,3,\ldots,K 
\label{cons} 
\end{eqnarray}
and thus parametrise a moduli space ${\cal N}(\Sigma)$ of complex
dimension $KN-N+1$ which is identified with the Coulomb branch of the
four-dimensional theory. 
\paragraph{}
The low-energy physics on the Coulomb branch is determined by
the periods of the differential $\lambda=vdz$ which is meromorphic on 
$\Sigma$. We choose a canonical basis $\{A^{I},B_{I}\}$ 
of homology one-cycles on $\Sigma$ obeying 
$A^{I}\cap B_{J}=\delta^{I}_{J}$ for $I,J=1,2,\ldots,r$ and define
corresponding periods,  
\begin{eqnarray}
a^{I} \,\,=\,\,\frac{1}{2\pi i}\oint_{A^{I}}\,\lambda & \qquad &
a^D_{I} 
\,\,=\,\,\frac{1}{2\pi i}
\frac{\partial {\cal F}}{\partial a_{I}}\,\,=\,\,
\frac{1}{2\pi i}\oint_{B_{I}}\,\lambda. \nonumber 
\end{eqnarray}
Here ${\cal F}(a)$ is a holomorphic prepotential which completely
determines the low-energy effective action. The bosonic part of the
action takes the form 
\begin{equation} \mathcal{L} = \frac{1}{4\pi}\im
  \tau_{IJ}\partial_m a^I\partial^m\bar{a}^J +
  \frac{1}{8\pi}\im \tau_{IJ}v_{m n}^I v^{J m n} +
  \frac{1}{8\pi}\re \tau_{IJ}
v_{m n}^I\tilde{v}^{J m n},\label{eqn:4daction}\end{equation}
where the
matrix of low-energy $U(1)^{r}$ 
gauge couplings is determined by the period matrix of $\Sigma$, 
\begin{equation} \tau_{IJ} = \frac{\partial^2 \F}{\partial
    a^I \partial a^J}.\label{eqn:SKmetric} 
\end{equation}
The scalar part of the effective action is a nonlinear $\sigma$-model
with target space ${\cal N}(\Sigma)$ and the corresponding
target space metric can be read from the scalar kinetic terms in 
(\ref{eqn:4daction});  
\[ ds^2 = \im \tau_{IJ}da^Id\bar{a}^J.\]
The existence of local coordinates $\{a^{I}\}$ in which the metric is
determined by a holomorphic prepotential in this way is the definition
of special K\"{a}hler geometry. In particular, the 
moduli space ${\cal N}(\Sigma)$ equipped with this metric is 
a special K\"{a}hler manifold. As the underlying quiver gauge theory
is superconformal, the Coulomb branch metric is also scale-invariant. 
This implies that the prepotential satisfies a further condition, 
\begin{equation} a^I\frac{\partial}{\partial a^I} \F = 2\F 
\label{eqn:conformalSK} \end{equation}
which will be important in the following.      
 \paragraph{}
The next step in the construction is to take the four-dimensional
quiver gauge theory described above and compactify it down to three
dimensions on a circle of radius $R$. The massless
fields in the resulting 
three-dimensional effective theory include the massless scalars $a^{I}$ of
the four-dimensional theory. We also obtain $r$ real scalars
$\theta_{e}^{I}$ from the Wilson lines of the massless $U(1)^{r}$ gauge
fields around the compact direction. A further $r$ real scalars 
$\theta_{m,I}$ arise after dualising the remaining components of the
$U(1)^{r}$ gauge field. The two sets of scalars are naturally periodic due to
invariance under large gauge transformations and the quantisation of
magnetic charge respectively. 
We normalise $\theta_{e}^{I}$ and $\theta_{m, I}$ to have
period $2\pi$ for all $I$. 
\paragraph{}
The massless scalars $\{a^{I},\theta_{e}^{I},\theta_{m, I}\}$ 
described above parametrise a Coulomb branch of
real dimension $4(KN-N+1)$ which we identify with the instanton
moduli space ${\cal M}$, where the compactification radius is
identified as\footnote{Note the apparent mismatch in dimensions in
  this identification which
  can be explained as follows. In the context of a four-dimensional
  gauge theory compactified on $S^{1}$, ${\cal M}$ corresponds to a
  vacuum moduli space and the natural coordinates are scalar fields 
which have mass dimension one-half in three dimensions. Here we are
  working instead in coordinates with dimensions of length as
  appropriate for the application to instantons living on the spacetime
  $\mathbb{R}^{2}\times T^{2}$.} $R=1/4\pi^{2}\ell^{2}$. 
The low-energy effective action is a
three-dimensional nonlinear $\sigma$-model which defines a metric on
this space. As the theory has eight supercharges the resulting metric 
must be hyper-K\"{a}hler.  It therefore has three linearly
  independent complex structures $J^{1}$,
$J^{2}$ and $J^{3}$. In fact there is a preferred complex
  structure \cite{Seiberg:Witten:1996}, $J=J^{3}$, 
which is independent of the compactification radius
  $R$. In this complex structure, the holomorphic coordinates can be
  taken as $\{a^{I},z_{I}\}$ where $a^{I}$ are the massless complex scalars of
  the four dimensional theory and,   
\[ z_I = \theta_{m, I} - \tau_{IJ}\theta_{e}^{J}\] 
parametrise an $r$-dimensional complex torus which can be identified
with the Jacobian, 
\[ {\cal J}(\Sigma) = 
\frac{\mathbb{C}^r}{\mathbb{Z}^r\oplus \tau \mathbb{Z}^r}.\]
In the preferred complex structure, the 
full Coulomb branch is therefore the total space of 
a fibre bundle\footnote{This statement is not quite precise due to an
  additional subtlety in the global definition of fibre coordinates known as
  the quadratic refinement \cite{Gaiotto:Moore:Neitzke:2008}. 
However, this will not play a role in the 
following.} $\mathcal{M}
\rightarrow \mathcal{N}(\Sigma)$ 
over the special K\"{a}hler base 
$\mathcal{N}(\Sigma)$ whose fibres are the Jacobian
tori ${\cal J}(\Sigma)$. With respect to a given complex structure a
hyper-K\"{a}hler manifold is a complex manifold with a holomorphic
symplectic form. In the local holomorphic coordinates introduced above
for $J=J^{3}$, the holomorphic symplectic form is given
as, 
\begin{eqnarray}
\eta & = & g.\left(J^{1}+iJ^{2}\right)=da^{I}\wedge dz_{I}. \nonumber 
\end{eqnarray}
\paragraph{}
The hyper-K\"{a}hler metric on ${\cal M}$ 
can be determined exactly
using the methods of  \cite{Gaiotto:Moore:Neitzke:2008}. 
Although complicated in general, when $R$ is much larger than any
other scales in the
problem the metric on the total space takes its {\em semi-flat} form: 
\begin{equation} G = R\im \tau_{IJ}da^Id\bar{a}^J +
  \frac{1}{4\pi^2R}\inv^{IJ}\delta z_I\delta 
\bar{z}_J,\label{eqn:Gsf}\end{equation}
where the one-form,  
\[\delta z_I = d\theta_{m, I}-\tau_{IJ}d\theta_{e}^J\]
is closed on the fibre $\mathcal{J}(\Sigma)$ but not on the total
space. The factor of $1/R$ in front of the second term in the metric 
means that that the volume of the fibre scales as $1/R^{r}$. 
In contrast, the factor $R$ in front of the first term is
conventional and can be removed by rescaling the coordinates $a^{I}$.  
\paragraph{}
The semi-flat metric is valid up to instanton 
corrections of order $\exp(-M_{\rm BPS}R)$
where $M_{\rm BPS}$ corresponds to the mass of a BPS state in the
four-dimensional quiver theory. To avoid later confusion we will
call these 3d instantons. The four-dimensional BPS mass formula is 
$M_{\rm  BPS}=|Z|$ where, 
\begin{eqnarray}
Z & =& n_{I}a^{I}+ m^{I}a^{D}_{I} \nonumber 
\end{eqnarray} 
and $n_{I}$ and $m^{I}$ are integer-valued electric and magnetic
charges for $I=1,2,\ldots,r$. 
Thus the semi-flat formula is valid in a limit where $R\rightarrow
\infty$ with $a^{I}$ held fixed away from the special submanifolds in
the moduli space where one or more of the cycles $\{A^{I},B_{I}\}$
degenerate. These are the famous Seiberg-Witten points of the 
four-dimensional theory where electric or magnetic BPS states become 
massless. The semi-flat metric itself has logarithmic singularities on these 
submanifolds which are smoothed out in the full metric by 3d instanton
corrections\footnote{Note however that 
singularities of higher codimension remain where Higgs or
mixed branches intersect the Coulomb branch.}.   
\paragraph{}
Together with the diagonal coupling $\tau_{cl}$, the couplings
$\tau_{j}$, for $i=1,2,\ldots,N$, 
control the weak-coupling expansion of the 4d ${\cal N}=2$
theory underlying the Coulomb branch description of
$\mathcal{M}$. Specifically the prepotential $\mathcal{F}(a)$ 
can be expanded in the limit $\tau_{cl}, \tau_{j}\rightarrow i\infty$ as, 
\begin{eqnarray}
\mathcal{F}(a) & = & \mathcal{F}_{\rm cl}\,+\,  \mathcal{F}_{\rm
  1-loop}\,+\, \mathcal{F}_{\rm inst} \nonumber     
\end{eqnarray}   
where the three terms correspond to classical, one-loop and instanton
contributions respectively in the 4d ${\cal N}=2$ theory. In the weak-coupling
region, natural coordinates on the Coulomb branch are provided by the
VEVs of the adjoint scalars $\Phi_{j}$
in the vector multiplet of the $SU(K)_{j}$ factor in the gauge group
$G$: 
\begin{eqnarray}
\langle \Phi_{j} \rangle & = & {\rm diag}\left(a_{j1}, a_{j2},\ldots,
  a_{jK}\right) \qquad{} {\rm for}\,
j=1,2,\ldots,N. \label{coords}
\end{eqnarray}
Here the coordinate index $I$ has been traded for an $SU(N)$ index, 
$i,j=1,2,\ldots,N$ and a $U(K)$ index $a,b=1,2,\ldots,K$. 
We can project out unwanted $U(1)$ factors by imposing the constraint, 
\begin{eqnarray}
\sum_{b=1}^{K}\, a_{j b} & = & \bar{a} \qquad{} {\rm for}\,
j=1,2,\ldots,N 
\label{cons2b}
\end{eqnarray}
where $\bar{a}$ is the scalar in the vector multiplet of the decoupled 
diagonal subgroup $U(1)_{D}$. In these coordinates we have, 
\begin{eqnarray}
{\cal F}_{\rm cl} &  = & \sum_{j=1}^{N}\, \sum_{b=1}^{K}\, 
\frac{1}{2}\tau_{j}a^{2}_{jb}
\label{cl} \\ 
{\cal F}_{\rm 1-loop} & = & 
-\frac{1}{2\pi i}\,\sum_{j=1}^{N}\, \sum_{a>b}^{K}\,f(a_{ja}-a_{jb})
+\frac{1}{4\pi i}\,\sum_{j=1}^{N}\, \sum_{a,b=1}^{K}\,f(a_{ja}-a_{j+1 b})
\label{1-loop}
\end{eqnarray}
where $f(x)=x^{2}(2\log x -1)$. In the 4d theory, the 
first term in (\ref{1-loop}) is
the one-loop contribution from integrating out the massive states of
the vector multiplets while the second comes from the bifundamental
hypermultiplets. Terms in the instanton contribution ${\cal F}_{\rm
  inst}$ are suppressed by powers of $\exp(2\pi i\tau_{\rm cl})$ and/or 
$\exp(2\pi i\tau_{j})$. We will call these effects, which do contribute to the
semi-flat metric, 4d instantons to distinguish them from
the 3d instantons discussed above which do not.   
\paragraph{} 
In particular, for ${\rm Im}\tau_{\rm cl}>>1$ and  ${\rm Im}\tau_{j}>>1$, 
we can approximate the prepotential ${\cal F}(a)$ 
by ${\cal F}_{\rm  cl}(a)$. The resulting moduli space
metric is flat. Taking account of the action of the Weyl group in each
$SU(K)$ factor and of the constraint (\ref{cons2}) the instanton moduli
space reduces to,   
\begin{eqnarray}
\mathcal{M}_{\rm cl} & = & \mathbb{R}^{2}\times T^{2}[\tau_{\rm cl}] \times 
\prod_{j=1}^{N}\, \frac{ {\rm Sym}^{K}\left(\mathbb{R}^{2}\times 
T^{2}[\tau_{j}]\right)}{\mathbb{R}^{2}\times 
T^{2}[\tau_{j}]}  
\nonumber 
\end{eqnarray}
with a flat metric. Here $T^{2}[\tau]$ is a two-dimensional
  torus of complex structure $\tau$ and 
${\rm Sym}^{K}$ denotes a $K$-fold symmetric product. The area of
each $T^{2}$ factor is proportional to $1/R\sim
\ell^{2}$.  
This form 
reflects a feature which is familiar from related studies of
instantons on tori (see especially \cite{Lee:Yi:1997}); 
$K$ instantons of gauge group $SU(N)$ fractionate
into $KN$ ``partons'' each of which carries a fraction of the total
instanton charge. The resulting configuration consists of $K$
identical partons of $N$ distinct species. 
A less familiar feature  is that the $N-1$ relative positions of the
centre of mass on $\mathbb{R}^{2}\times T^{2}$ 
of each of the $N$ distinct species of parton are frozen.   
\paragraph{}
The one-loop correction coming from ${\cal F}_{\rm
  1-loop}$ introduces logarithmic singularities in the semi-flat
metric. In particular singularities occur where electrically charged
particles from the vector and hyper-multiplets become light. The
inclusion of instantons corrects the picture further, resulting
in loci where magnetically charged states become light. 
 
\subsection{The Higgs branch description}
\paragraph{}
The Higgs branch description of the hyper-K\"{a}hler manifold ${\cal
  M}$ starts from the ADHM-Nahm transform for $K$ $SU(N)$ instantons
on $\mathbb{R}^{2}\times T^{2}$ with Wilson lines 
\cite{Kapustin:Sethi:1998}. 
This results in a $U(K)$ gauge theory on the dual torus $\hat{T}^{2}$
which, up to a rescaling of the area, 
we identify with the canonical torus $E(\tau)$. 
The resulting model 
has localised impurities at the $N$ points $z=Z_{i}$ specified by the
Wilson lines. The theory contains a $U(K)$ gauge field with components
$A_{z}(z,\bar{z})$ and $A_{\bar{z}}(z,\bar{z})$ as well as a complex
scalar field $\phi(z,\bar{z})$ in the adjoint of the gauge group. 
It also contains localised impurities $\{Q_{i},\tilde{Q}_{i}\}$ 
in the ${\bf K}\oplus\bar{\bf K}$ of $U(K)$ defined at the marked
points $z=Z_{i}$ on $\hat{T}^{2}$ for $i=1,2,\ldots,N$. In general, we
construct the Higgs branch by imposing the F- and D-term equations of
the theory and modding out by the $U(K)$ gauge symmetry. In order to
describe ${\cal M}$ as a complex symplectic manifold (in the
preferred complex structure $J$) it suffices instead 
to impose only the F-term
equation which reads, 
\begin{eqnarray}
\partial_{\bar{z}}\phi \,+\, \left[A_{\bar{z}}, \phi\right] & = & 
2\pi i\, \sum_{i=1}^{N}\,Q_{i}\tilde{Q}_{i}\, 
\delta^{(2)}\left(z-Z_{i}\right) \label{Fterm} 
\end{eqnarray}
and divide out by the complexified gauge symmetry \cite{Luty:Taylor:1995}, 
\begin{eqnarray}
A_{z} & \rightarrow & \Omega A_{z}\Omega^{-1}\,-\, 
\Omega\partial_{z}\Omega^{-1} \nonumber \\
\phi & \rightarrow & \Omega\phi\Omega^{-1}\qquad{}
\Omega\left(z,\bar{z}\right)\in GL\left(K, \mathbb{C}\right).
\label{cgt}
\end{eqnarray}
Following the analysis of \cite{Nekrasov:1995,ND:Hollowood:Kumar:2001}, we first use the off-diagonal
part of the complex gauge symmetry to bring $A_{\bar{z}}$ to the
diagonal form, 
\begin{eqnarray}
A_{\bar{z}} & = & \frac{\pi i}{2(\bar{\omega}_{2}\omega_{1}-\bar{\omega}_{1}
\omega_{2})}\,{\rm diag}\left(X_{1}, X_{2},\ldots,X_{K}\right). 
\nonumber 
\end{eqnarray}
The remaining degrees of freedom $X_{a}$ have a natural periodic
identification, 
\begin{eqnarray}
X_{a} & = & X_{a}+2n\omega_{1}+2m\omega_{2} \qquad{} \qquad m, n\in
\mathbb{Z} 
\nonumber 
\end{eqnarray}
corresponding to large gauge transformations. In this gauge we may
solve (\ref{Fterm}) explicitly for $\phi(z,\bar{z})$. For the diagonal
elements we find, 
\begin{eqnarray}
\phi_{aa}(z) & = & P_{a} \,+\, \sum_{i=1}^{N}\,Q^{ia}
\tilde{Q}^{ia}\, \xi\left(z-Z_{i}\right) \nonumber 
\end{eqnarray}
where $P_{a}$, for $a=1,2,\ldots,K$, are undetermined complex
constants. The off-diagonal elements are given as, 
\begin{eqnarray}
\phi_{ab}(z) & = & \sum_{i=1}^{N} \,Q^{ia}\tilde{Q}^{ib} \,
\frac{\sigma\left(X_{ab}+z-Z_{i}\right)}{\sigma\left(X_{ab}\right)\sigma
\left(z-Z_{i}\right)} \nonumber
\end{eqnarray}   
for $a, b=1,2,\ldots,K$ with $a\neq b$ where $X_{ab}=X_{a}-X_{b}$ and
$\sigma(z)$ is the Weierstrass $\sigma$-function. 
\paragraph{}
The resulting description of the complex manifold ${\cal M}$ is
given in holomorphic coordinates provided by the $2KN+2K$ complex
parameters $\{Q^{ia}, \tilde{Q}^{ia},X^{a},P^{a}\}$ subject to the
$K+N-1$ constraints\footnote{Here the first set of $N-1$ independent 
constraints in (\ref{cons2}) and their accompanying gauge transformations 
in (\ref{cgt2}) correspond to the restriction from $U(K)$ to $SU(K)$ in each
factor of the Coulomb branch gauge group $G$. The second set of
$K$ constraints are implied by the diagonal components of equation 
(\ref{Fterm}), while the accompanying 
gauge transformations correspond to the diagonal
generators of $GL(K,\mathbb{C})$.}, 
\begin{eqnarray}
\sum_{a=1}^{K}\,Q^{ia}\tilde{Q}^{ia} = 0 & \qquad{} & 
\sum_{i=1}^{N}\,Q^{ia}\tilde{Q}^{ia} =0
\label{cons2}
\end{eqnarray}
and the residual gauge symmetry, 
\begin{eqnarray}
Q_{ia} & \rightarrow & \xi_{a}\,Q_{ia}\,\xi_{i} \nonumber \\ 
\tilde{Q}_{ia} & \rightarrow & \xi^{-1}_{a}\,\tilde{Q}_{ia}\,\xi^{-1}_{i} \qquad{}
\xi_{a}, \xi_{i}\in \mathbb{C} \label{cgt2} 
\end{eqnarray}
of dimension $K+N-1$ thus giving a complex manifold of the expected
complex dimension $2(KN-N+1)$. 
\paragraph{}
The above construction is an infinite-dimensional version of the
standard hyper-K\"{a}hler quotient. 
In the preferred complex structure it provides a quotient construction
of the complex symplectic manifold $({\cal M},J,\eta)$. In particular, one
starts from the infinite-dimensional space $\mathcal{M}_{\infty}$, 
spanned by the fields
of the 2d theory $\{\phi(z,\bar{z}), A_{z}(z,\bar{z}),
Q^{i},\tilde{Q}^{i}\}$ endowed with the holomorphic symplectic form, 
\begin{eqnarray}
\eta_{\infty} & = & \,\int d^{2}z\,\sum_{a,b=1}^{K}\, 
d\phi_{ab}(z,\bar{z})\wedge dA_{ba}^{z}(z,\bar{z}) 
\,+\,\sum_{i=1}^{N}
\sum_{a=1}^{K}\, dQ^{ai}\wedge d\tilde{Q}^{ai} .
\nonumber 
\end{eqnarray}
After imposing the F-term (\ref{Fterm}) equation, the symplectic form 
descends to the quotient space where it takes the form, 
\begin{eqnarray}
\eta & = & \sum_{a=1}^{K}\, dX^{a}\wedge dP^{a} \,+\, \sum_{i=1}^{N}
\sum_{a=1}^{K}\, dQ^{ia}\wedge d\tilde{Q}^{ia}. \nonumber 
\end{eqnarray}
Finally the constraints (\ref{cons2}) can be imposed and the diagonal
complex gauge transformations (\ref{cgt2}) divided out in a further
finite-dimensional symplectic quotient. 
\paragraph{}
The symplectic quotient construction also provides a tower of Poisson-commuting
Hamiltonians which promote $({\cal M},J,\eta )$ to an
algebraic integrable system. Here we define the Poisson bracket of any
two holomorphic functions $f$ and $g$ on ${\cal M}$ as, 
\begin{eqnarray}
\{f,g\}= \left(\eta^{-1}\right)^{uv}\,\frac{\partial f}{\partial
  w^{u}}\frac{\partial g}{\partial w^{v}}
\nonumber 
\end{eqnarray}
where $w^{u}$, $u=1,2,\ldots,2r$ are holomorphic coordinates on
$\mathcal{M}$. 
The quotient construction guarantees that all quantities which are invariant
under complex gauge transformation (\ref{cgt}) and which Poisson-commute
on the big phase space $\mathcal{M}_{\infty}$ 
will descend to Poisson-commuting quantities on
the quotient space. In particular we have, 
\begin{eqnarray}
\left\{\, {\rm Tr}_{K} \phi^{l}(z),\,  {\rm Tr}_{K} \phi^{m}(z')\right\}
& = & 0 \qquad{} \forall\, l,m\in\mathbb{Z} \qquad{}\forall\, z, z'\in
E(\tau). \nonumber 
\end{eqnarray}
The tower of conserved charges is encoded in the {\em spectral curve} of the
integrable system, 
\begin{eqnarray}
F(v,z) & = & {\rm det}\left(v \mathbb{I}_{K} \,-\,
\phi(z)\right) \,\,=\,\,0.
\end{eqnarray}
Examination of the singularities of $F(v,z)$ reveals that it
obeys the same constraints as the function of the same name appearing in
(\ref{curve}). Thus the spectral curve is precisely the same as the complex 
curve $\Sigma$ which appears in the Coulomb branch description of the
moduli space. Hence using the parametrisation (\ref{param}) we find
that the quantities $\{{\cal H}^{(i)}_{a}$,  ${\cal H}^{(0)}_{a}\}$
subject to the constraint (\ref{cons}) provide a tower of $KN-N+1$
commuting Hamiltonians on the complex phase space $({\cal
  M},J,\eta)$.    
\paragraph{}
When expressed in terms of the Higgs branch coordinates, $\{Q^{ia},
\tilde{Q}^{ia},X^{a},P^{a}\}$, 
the integrable system takes the form of a holomorphic 
many-body problem \cite{Nekrasov:1995}. In particular, the
symplectic form yields Poisson brackets, 
\begin{eqnarray}
\{X_{a},P_{b}\}\,=\,\delta_{ab} & \qquad{} & \{S^{a}_{ij},
S^{b}_{kl}\}\,=\,
\delta_{ab}\left(\delta_{jk}S^{a}_{il}\,-\,\delta_{il}S^{a}_{kj}
\right) 
\label{sln} 
\end{eqnarray} 
where we have defined spin variables, 
\begin{eqnarray}
S^{a}_{ij} & = &  Q^{a}_{i}\tilde{Q}^{a}_{j}
\end{eqnarray}
for $a=1,2,\ldots, K$ which obey the $SL(N,\mathbb{C})$ Poisson algebra as given
in (\ref{sln}). 
Explicit expressions for the Hamiltonians can be found in 
\cite{ND:Hollowood:Kumar:2001}. 
In the simplest case where all $K$ punctures coincide, $Z_{i}=Z_{j}$
for all $i,j=1,2,\ldots,N$, one particular quadratic combination 
of the Hamiltonians takes the form, 
\begin{eqnarray}
H_{2} & = & \sum_{a=1}^{K} P_{a}^{2}\,+\,\sum_{a\neq b}
\sum_{i,j=1}^{N} 
\, S^{a}_{ij}S^{b}_{ji}\, {\cal P}\left(X_{a}-X_{b}\right)
\end{eqnarray}
where ${\cal P}(z)$ is the Weierstrass elliptic function. This is the
Hamiltonian for a generalisation of the elliptic Calogero-Moser
model in which each of the $K$ particles carries an $SL(N,\mathbb{C})$
spin. The integrabilty of this model was used to solve it exactly in 
\cite{Kricheveretal:1994}. 
\paragraph{}
Another interesting case is the limit $\tau\rightarrow i\infty$, with
$\tau_{j}$ held fixed, where the elliptic quiver degenerates to a
linear quiver based on the Dynkin diagram of $A_{N-1}$. In the
simplest case $N=2$, the resulting theory is ${\cal N}=2$ SUSY QCD with
gauge group $U(K)$ and $2K$ massless hypermultiplets in the fundamental
representation. In this limit, the variables $P_{a}$
describing the momenta of the Calogero particles are frozen and the 
conjugate variables $X_{a}$ are gauged away.  
The resulting Hamiltonians for the spins $S^{a}_{ij}$ describe a
classical Heisenberg spin chain with $SL(N,\mathbb{C})$ spins at $K$
sites \cite{Gorskyetal:1996,Gorsky:Gukov:Mironov:1997}. 
The classical Casimirs of the $SL(N,\mathbb{C})$ Poisson algebra (\ref{sln})
vanish at each site and the off-diagonal couplings $\tau_{j}$
correspond to twisted boundary conditions for the chain. 
\paragraph{}
The integrable system provides a nice way of understanding the
equivalence between the Higgs and Coulomb branch descriptions of
$({\cal M},J,\eta)$. As we have already noted above, the spectral curve
of the integrable system coincides with the Seiberg-Witten curve
$\Sigma$ of the Coulomb branch description. The commuting Hamiltonians
of the integrable system coincide with the moduli of the curve which
parametrise the base of the Jacobian fibration in the Coulomb branch
picture. We then recognise the coordinates $\{a^{I},z_{I}\}$ as the
action-angle variables of the system. Indeed, the solution to the
model given in \cite{Kricheveretal:1994} precisely takes the form 
of a linear flow on the Jacobian $\mathcal{J}(\Sigma)$.   

\section{The $\sigma$-model}
\paragraph{}
We now wish to consider a quantum mechanical $\sigma$-model whose
target space is the instanton moduli space ${\cal M}$ described in the
previous section with the semi-flat metric (\ref{eqn:Gsf}). We work
in local holomorphic 
coordinates $\{a^{I},z_{I}\}$, with $I=1,2,\ldots,r$ where, as above, 
\[ z_I = \theta_{m\,I} - \tau_{IJ}\theta^{J}_{e}.\] 
As the target space is hyper-K\"{a}hler, the bosonic $\sigma$-model
action has a completion with ${\cal N}=(4,4)$ supersymmetry 
\cite{AlvarezGaume:Freedman:1981} and a $USp(4)\simeq SO(5)$
R-symmetry \cite{FOF:Kohl:Spence:1997,Verbitsky:1990}. 
\paragraph{}
The resulting supersymmetric 
$\sigma$-model contains fermions\footnote{In this section we will adopt
  the standard convention that all anti-holomorphic coordinate indices are
  barred.} $\psi^{I A}$, 
$\bar{\psi}^{\bar{I}\bar{A}}$ in the ${\bf 4}$ and ${\bar{\bf 4}}$
of $USp(4)$ respectively in addition to the bosonic coordinates
$a^{I}$, $z_{I}$. The couplings in the $\sigma$-model 
Lagrangian are completely determined by the
holomorphic prepotential ${\cal F}(a)$ introduced above. In the following, 
${\cal F}^{(3)}_{IJK}$ and ${\cal F}^{(4)}_{IJKL}$ denote the third
and fourth mixed partial derivatives of the prepotential 
with respect to the base coordinates $a^{I}$. 
The Lagrangian reads \cite{ND:AS:2014},  
\begin{align} \label{lag} \begin{aligned} \mathcal{L}_{\sigma} 
&= g_{I\bar{J}}\dot{a}^I\dot{\bar{a}}^{\bar{J}} + 
g^{I\bar{J}}\frac{\delta z_I}{\delta t}
\frac{\delta \bar{z}_{\bar{J}}}{\delta t} \\ &+
 ig_{I\bar{J}}\bar{\psi}^{\bar{J}}_A D_t \psi^{IA}
 -\frac{1}{12}\re\left[\epsilon_{ABCD}G_{IJKL}
\psi^{IA}\psi^{JB}\psi^{KC}\psi^{LD}\right] \\ &- 
\frac{1}{2}R_{I\bar{J}K\bar{L}}\psi^{IA}\bar{\psi}^{\bar{J}}_A
\psi^{KB}\bar{\psi}^{\bar{L}}_B. \end{aligned} \end{align}
where, 
\begin{align} \begin{aligned} 
D_t\psi^{IA} &= \dot{\psi}^{IA} 
-\frac{i}{2}\inv^{IL}\F^{(3)}_{LJK}\dot{a}^J\psi^{KA} \\
\frac{\delta z_I}{\delta t} &= 
 \dot{z}_{I}-\F^{(3)}_{IJK}\inv^{KL}
\im z_L \dot{a}^J - \frac{1}{4}\F^{(3)}_{IJK}\Omega_{AB}
\psi^{JA}\psi^{KB} \end{aligned} \label{eqn:variablechange} 
\end{align}
with $\Omega_{AB}$ being the invariant antisymmetric tensor of
$USp(4)$, and,  
\begin{equation}R_{I\bar{J}K\bar{L}} = -\frac{1}{4}\inv^{M\bar{N}}
\F^{(3)}_{IKM}\bar{\F}^{(3)}_{\bar{J}\bar{L}\bar{N}}
\label{eqn:baseRiemann} \end{equation}
denotes the base space Riemann tensor. We also define the 
totally symmetric base space tensor,
\begin{align} \begin{aligned} G_{IJKL} &= & -\frac{i}{2}\F^{(4)}_{IJKL} +
 \frac{1}{4} \inv^{MN}\left(\F^{(3)}_{ILM}\F^{(3)}_{JKN}
 + \F^{(3)}_{JLM}\F^{(3)}_{IKN} + 
\F^{(3)}_{KLM}\F^{(3)}_{IJN}\right). \end{aligned}
 \label{eqn:chiraltensor}\end{align}
\paragraph{}
The Hamiltonian formulation of the $\sigma$-model was worked out in
detail in \cite{ND:AS:2014}. The dynamical variables are the base coordinates
$a^{I}$ and their ``covariant'' conjugate momenta 
$\Pi_{I}={\rm Im}\tau_{I\bar{J}}\dot{\bar{a}}^{\bar{J}}$, the fibre
coordinates $z_{I}$ and their canonical conjugate momenta $P^{I}=
{\delta{\cal L}_{\sigma}}/\delta \dot{z}_{I}$ as well as the fermions $\psi^{I A}$
and $\bar{\psi}^{\bar{I}\bar{A}}$. The relevant non-zero (anti-)commutation
relations are,        
\begin{subequations} \nonumber \label{eqn:Diracbrackets} \begin{align}
    \left[a^I,\Pi_J\right] &= i\delta^I_J\\ \left[z_{I},\Pi_J\right]
    &= i\im z_K \F^{(3)}_{IJL}\inv^{KL} & \left[z_{I},P^{J}\right]
    &= i\delta_{I}^{J}  \\ \left[\Pi_I,P^{J}\right] &=
    \frac{1}{2}\F^{(3)}_{IKL}\inv^{JL}P^{K} &
    \left[\Pi_I,\bar{P}^{\bar{J}}\right] &=
    -\frac{1}{2}\F^{(3)}_{IKL}\inv^{\bar{J}L}P^{K} 
 \end{align} 
\end{subequations}
and 
\[ \left[\Pi_I,\psi^{JA}\right] = i\Gamma^J_{IK}\psi^{KA}, \quad
\left[\Pi_I,\bar{\psi}^{\bar{J}\bar{A}}\right] = 
0\]
\begin{equation} \nonumber
\left\{\psi^{IA},\bar{\psi}^{\bar{J}\bar{B}}\right\} =
    \delta^{A\bar{B}}\inv^{I\bar{J}}.
\end{equation}
\paragraph{}
In terms of these variables, the $\sigma$-model Hamiltonian
reads, 
\begin{align} \nonumber \begin{aligned} H &= \inv^{I\bar{J}}\Pi_I
\bar{\Pi}_{\bar{J}}
+ 
\frac{1}{2}R_{I\bar{J}K\bar{L}}\psi^{IA}\bar{\psi}^{\bar{J}}_A\psi^{KB}
\bar{\psi}^{\bar{L}}_B
\\ &+ \frac{1}{12}\re 
\left(\epsilon_{ABCD}G_{IJKL}\psi^{IA}\psi^{JB}\psi^{KC}\psi^{LD}\right)
\\ &+ \im \tau_{I\bar{J}}P^{I}\bar{P}^{\bar{J}}
+ \frac{1}{2}\re \left( \F^{(3)}_{IJK}\Omega_{AB}\psi^{JA}\psi^{KB}P^{I}\right)
\end{aligned} \end{align}
and the supercharges, 
\begin{align} \nonumber \begin{aligned} Q^{A} &= \psi^{IA}\Pi_I
+ \frac{1}{12}\epsilon^A_{~\bar{B}\bar{C}\bar{D}} \bar{\F}^{(3)}_{\bar{I}\bar{J}\bar{K}}
\bar{\psi}^{\bar{I}\bar{B}}\bar{\psi}^{\bar{J}\bar{C}}\bar{\psi}^{\bar{K}\bar{D}}
\\ &+ \im \tau_{I\bar{J}}P^{I}\Omega^A_{~\bar{B}}\bar{\psi}^{\bar{J}\bar{B}}
\end{aligned} \end{align}
and $\bar{Q}^{\bar{A}}=(Q^{A})^{\dagger}$ transform 
in the ${\bf 4}\oplus\bar{\bf 4}$ of $USp(4)$,  
commute with $H$, and obey the supersymmetry algebra,
\begin{eqnarray}
\{Q^{A}, Q^{\bar{B}}\} & = & \delta^{A\bar{B}} \, H \nonumber \\ 
\{Q^{A}, Q^{B}\} & = & \{Q^{\bar{A}}, Q^{\bar{B}}\}\, = \, 0. \nonumber 
\end{eqnarray}
\paragraph{}
A striking feature of the
semi-flat $\sigma$-model Lagrangian (\ref{lag}) is the presence of $2(KN-N+1)$ 
$U(1)$ global symmetries corresponding to constant shifts of the real
fibre coordinates $\theta_{e}$, $\theta_{m}$. In the Hamiltonian
formalism these lead to the Noether charges, 
\begin{eqnarray}
Q_{e\,I}\,=\, \tau_{IJ}P^{J}+  \bar{\tau}_{IJ}\bar{P}^{J}& \qquad{} & 
Q^{I}_{m}  \,=\, P^{I}+\bar{P}^{I} .
\label{charges}
\end{eqnarray}
One may readily check that these charges are conserved and also 
commute with each other. 
Although the charges are related to the conserved quantities
of the integrable system described in the previous section it is
important to note that the full $\sigma$-model is certainly not
integrable.  The number of real conserved
charges is $2r=2(KN-N+1)$ which is half the dimension of the target
space and therefore half the number needed for
integrability. However it will be useful to label states in the
$\sigma$-model by the corresponding eigenvalues of these conserved
quantities.       
\paragraph{}
In particular, we will start by focusing 
on states carrying zero momentum in the fibre directions, 
$Q_{e}=Q_{m}=0$, or equivalently $P=0$. 
This sector of the theory has an enhanced conformal 
symmetry in which the Hamiltonian $H$ is joined by a
dilatation operator and special conformal generator,   
\begin{align} \nonumber \begin{aligned} 
 D &= a^I\Pi_I + \bar{a}^{\bar{I}}\bar{\Pi}_{\bar{I}}
 \\ K &= \im \left(\frac{\partial \F}{\partial a^I}\bar{a}^I\right)
\end{aligned} \end{align}
which form an $SL(2,\mathbb{R})\simeq SO(2,1)$ conformal algebra, 
\begin{subequations} \nonumber \begin{align} \left[H,K\right]
 &= -iD \qquad \left[D,K\right] = -2iK \qquad \left[D,H\right] = 2iH.
 \end{align}
 \end{subequations}
Further the $SO(5)$ R-symmetry of the full $\sigma$-model is enhanced
to $U(4)\simeq U(1)\times SU(4)$ with generators, 
\begin{align} \label{Rsym} \begin{aligned} 
\mathcal{R} &= i\left(a^I\Pi_I - \bar{a}^{\bar{I}}\bar{\Pi}_{\bar{I}}\right)
 + \frac{1}{2}\im \tau_{I\bar{J}}\psi^{IA}\bar{\psi}^{\bar{J}}_A
 &&\\ R^{A\bar{B}} &=
 i\im \tau_{I\bar{J}}\left(\psi^{IA}\bar{\psi}^{\bar{J}\bar{B}}
-\frac{1}{4}\delta^{A\bar{B}}\psi^{IC}\bar{\psi}^{\bar{J}}_C\right).
\end{aligned}
\end{align}
The supercharges $Q^{A}$ and 
$\bar{Q}^{\bar{A}}$ which now transform in
the ${\bf 4}\oplus\bar{\bf 4}$ of $U(4)$ are supplemented
by the superconformal generators,   
\begin{align} \nonumber \begin{aligned} 
S^A &= \im \tau_{I\bar{J}}\bar{a}^{\bar{J}}\psi^{IA}
 & \qquad{} \bar{S}^{\bar{A}} &= \left(S^A\right)^{\dagger} 
\end{aligned} \end{align}
also in the  ${\bf 4}\oplus\bar{\bf 4}$ of $U(4)$. 
The generators listed above close onto an ${\bf su}(1,1|4)$
superalgebra of conformal transformations \cite{ND:AS:2014}. This is precisely
the superconformal symmetry algebra of ${\cal N}=4$ supersymmetric
Yang-Mills compactified on a null circle. 
\paragraph{}
The spectrum of states with zero momentum in the fibre directions can
be decomposed in irreducible representations of the light-cone
superconformal symmetry $SU(1,1|4)$. In the context of the application
to DLCQ, an obvious goal for our analysis
is to try to determine this spectrum or equivalently determine the spectrum of
the dilatation operator $D$. 
For many purposes it is convenient
to perform a similarity transformation to a different basis of
generators for the superconformal group 
\cite{Nishida:Son:2007,Kim:Kim:Koh:Lee:Lee:2011}. Let us consider the following
family of generators labelled by a parameter $\mu$ with
the dimensions of mass: 
\begin{eqnarray}
\mathcal{L}_{0}[\mu] & = & \frac{1}{\mu}H+\mu K \nonumber \\ 
\mathcal{L}_{\pm}[\mu] & = & \frac{1}{2}
\left(\frac{1}{\mu}H-\mu K \pm i D\right) \nonumber 
\end{eqnarray} 
which obey the ${\bf sl}(2,\mathbb{R})$ commutation rules in the
standard form, 
\begin{eqnarray}
\left[\mathcal{L}_{0}, \mathcal{L}_{\pm}\right]\,=\,\pm
2\mathcal{L}_{\pm}  & \qquad{} &  \left[\mathcal{L}_{+}, 
\mathcal{L}_{-}\right]\,=\, -2\mathcal{L}_{0}. 
\nonumber
\end{eqnarray}
Under the similarity transformation, 
the problem of diagonalising $D$ can be mapped onto the equivalent
problem of diagonalising $\mathcal{L}_{0}$; 
\begin{eqnarray}
{\rm Spec}\left(D\right) & = & {\rm Spec}\left(\mathcal{L}_{0}[\mu]\right)  
\qquad{} \forall \mu \in \mathbb{R}/\{0\} .
\nonumber
\end{eqnarray}
More explicitly we have, 
\begin{eqnarray}
\mathcal{L}_{0}[\mu] & = & \frac{1}{\mu}\inv^{IJ}\Pi_I
\bar{\Pi}_{J} \,+\, \mu  \im \tau_{IJ} a^{I}\bar{a}^{J}\,\,+\,\,{\rm fermions}.
\nonumber 
\end{eqnarray}
Thus we see that the special conformal generator $K$ contributes a
harmonic potential centered at the origin $a^{I}=0$. It follows that
eigenstates of ${\cal L}_{0}$ are supported near this point which, 
unfortunately, is highly singular. Indeed all the
singular submanifolds where one or more cycles of $\Sigma$ degenerate
intersect at the origin. To make further progress we either need to find a
way of resolving the singularity or restrict our attention
to states whose wavefunctions are supported away from the
singularities. In fact we will focus on the latter option.  
\paragraph{}
We now consider generic states with non-zero momenta along the fibre
directions. Na\"{i}vely the scale set by the volume of the fibre breaks
superconformal invariance. Nevertheless we find that the 
operator $\mathcal{L}_{0}$ has a non-trivial ``lift'' to the full
theory. We define the operator, 
\begin{eqnarray}
\mathbb{H}[\mu] & = & \frac{1}{\mu}H +\mu K-
\im \tau_{IJ} P^{I}\bar{a}^{J}- \im \tau_{IJ}\bar{P}^{I}a^{J} 
\nonumber 
\end{eqnarray}
which reduces to the $SL(2,\mathbb{R})$ conformal generator 
$\mathcal{L}_{0}[\mu]$ when the fibre momentum vanishes. 
In fact, for each value of $\mu$ we find that $\mathbb{H}[\mu]$
is part of a non-trivial $OSp(1,1|4)$ subgroup of the $SU(1,1|4)$ 
superconformal invariance which survives in the full theory. 
To exhibit this, we define supercharges,
\begin{eqnarray} 
\mathbb{Q}^{A}_{+}\,=\,
q^{A}_{+}+i\Omega^{A}_{\,\bar{B}}\bar{q}^{\bar{B}}_{+} & \qquad{} & 
\mathbb{Q}^{\bar{A}}_{-}\,=\,
\bar{q}^{\bar{A}}_{-}-i\Omega^{\bar{A}}_{\, B}{q}^{B}_{-} \nonumber 
\end{eqnarray}
where,  
\begin{eqnarray}
q^{A}_{\pm} \, = \,\frac{1}{\sqrt{\mu}}Q^{A}\pm i\sqrt{\mu}S^{A}
& \qquad{} & \bar{q}^{\bar{A}}_{\pm} \,=\,  
\frac{1}{\sqrt{\mu}}\bar{Q}^{\bar{A}}\pm i\sqrt{\mu}\bar{S}^{\bar{A}}
\nonumber 
\end{eqnarray}
and $USp(4)$ R-symmetry generators, 
\begin{eqnarray}
\mathbb{R}^{A\bar{A}} & = & R^{A\bar{A}} -\Omega^{A}_{\,\bar{B}}
\Omega^{\bar{A}}_{\,B}R^{B\bar{B}}.
\nonumber 
\end{eqnarray}
These generators form an $OSp(1,1|4)$ super-algebra with non-vanishing
(anti-)commutators, 
\begin{eqnarray}
\left[\mathbb{H}, \mathbb{Q}^{A}_{+}\right]\,=\, \mathbb{Q}^{A}_{+} 
& \qquad{} & \left[\mathbb{H}, \mathbb{Q}^{\bar{A}}_{-}\right]\,=\, -
\mathbb{Q}^{\bar{A}}_{-} \nonumber 
\end{eqnarray} 
and
\begin{eqnarray}
\left\{  \mathbb{Q}^{A}_{+}, \mathbb{Q}^{\bar{A}}_{-}\right\} 
& = & 2\,\delta^{A\bar{A}}\mathbb{H}\,+\, 2i\,\mathbb{R}^{A\bar{A}}.
\nonumber 
\end{eqnarray}
We omit the commutators for $\mathbb{R}^{A\bar{A}}$ with the other generators as
these are dictated by assignment of $\mathbb{Q}_{+}$ and
$\mathbb{Q}_{-}$ to $USp(4)$ representations ${\bf 4}$ and $\bar{\bf
  4}$ respectively. Despite the fact that the $SL(2,\mathbb{R})$
invariance of the zero momentum sector is broken, the surviving
subgroup preserves some of the important features of superconformal
invariance. In particular, the supercharges do not commute with
$\mathbb{H}$, instead they have eigenvalues $\pm 1$ under commutation
with this operator. In addition, the $USp(4)$ R-symmetry arises in the
commutators of the supercharges and is therefore part of the algebra
rather than an outer automorphism.    
\paragraph{}
To further motivate the definition of the new operator we consider its
form in more detail,  
\begin{eqnarray}
\mathbb{H}[\mu] & = & \frac{1}{\mu}\inv^{IJ}\Pi_I
\bar{\Pi}_{J} \,+\, \frac{1}{\mu}  \im \tau_{IJ}\left(P^{I}- \mu a^{I}\right)
\left(\bar{P}^{J}-\mu \bar{a}^{J}\right)\,\,+\,\,{\rm fermions}.
\label{new} 
\end{eqnarray}
Hence we see that the harmonic potential contributing to
$\mathcal{L}_{0}$ has been translated and now has its centre at, 
\begin{eqnarray}
a^{I} & = & \frac{1}{\mu}\, P^{I} \nonumber 
\end{eqnarray}
for $I=1,2,\ldots,r$, and it follows that eigenfunctions of 
$\mathbb{H}$ are exponentially localised near
this point. As the $P^{I}$ (or more precisely the real quantities
$Q_{e}$ and $Q_{m}$ defined above) are conserved quantities,
we can set them to arbitrary fixed values (at least classically) and the
condition picks out a single point on the base. For such generic values,
the metric on $\mathcal{M}$ is non-singular. Hence we expect that
the operator $\mathbb{H}$ has a discrete spectrum with non-singular
eigenfunctions.      
\paragraph{}
What is the physical interpretation of $\mathbb{H}$? First, note that,
up to a rescaling, $\mathbb{H}$ differs from the $\sigma$-model
Hamiltonian H only by a shift of the canonical fibre
momentum\footnote{To check this statement for the fermionic sector of
  the model one needs to use the scale invariance condition
  (\ref{eqn:conformalSK}) for the prepotential.},
$P^{I} \rightarrow P^{I}\,-\,\mu a^{I}$;  
\begin{eqnarray}
\mathbb{H} & = & \left.\frac{1}{\mu}H\right|_{P^{I} \rightarrow P^{I}\,-\,\mu a^{I}}.
\nonumber 
\end{eqnarray}
As the original model describes a free particle moving on the instanton
moduli space $\mathcal{M}$, the modification has a natural
interpretation as coupling to a background vector potential specified
by the one-form, 
\begin{eqnarray}
\mathcal{A} & = & \mu a^{I}dz_{I}\,+\,\mu \bar{a}^{I}d\bar{z}_{I}
\nonumber 
\end{eqnarray} 
which corresponds to a target space magnetic field, 
\begin{eqnarray}
\mathcal{F} & = & \mu da^{I}\wedge dz_{I} \,+\,\mu d\bar{a}^{I}\wedge
d\bar{z}_{I}\, =\,\mu\,\left(\eta\,+\,\bar{\eta}\right)
\nonumber 
\end{eqnarray}
proportional to (the real part of) the globally-defined 
holomorphic symplectic form $\eta$.  
Equivalently we can pass back to the Lagrangian formulation and work with a new
Lagrangian, 
\begin{eqnarray}
\tilde{{\cal L}}_{\sigma} & = & {\cal L}_{\sigma}\,+\, 
\delta{\mathcal L} \nonumber \\ 
& = & {\cal L}_{\sigma} \,-\,
\mu a^{I}\dot{z}_{I}\,-\,\mu \bar{a}^{I}\dot{\bar{z}}_{I}
\nonumber 
\end{eqnarray}

\section{Discrete light-cone quantisation}
\paragraph{}
Finally we are ready to discuss the main physical application for the
quantum mechanical $\sigma$-model developed above: the description of
${\cal N}=4$ supersymmetric Yang-Mills theory in discrete light-cone
quantisation (DLCQ). More precisely we will consider the
six-dimensional $(2,0)$ superconformal field theory of Type $A_{N-1}$
compactified to four dimensions on a torus $T^{2}$ of complex
structure $\tau_{\rm cl}$ and area $\mathcal{A}=4\pi^{2}{\rm
  Im}\tau_{\rm cl}\,\ell^{2}$. The compactified theory flows in the IR to
${\cal N}=4$ SUSY Yang-Mills with gauge group $SU(N)$ and complexified 
coupling, 
\begin{eqnarray}
\tau_{\rm cl} & = & \frac{4\pi i}{g^{2}}\,+\, \frac{\theta}{2\pi}.
\nonumber
\end{eqnarray}
The ${\cal N}=4$ theory is defined on
four-dimensional Minkowski space $\mathbb{R}^{1,3}$ with coordinates 
$\{x_{0},x_{1},x_{2},x_{3}\}$. In addition to the degrees of freedom
of ${\cal N}=4$ SUSY Yang-Mills, the full theory contains two towers
of Kaluza-Klein states, with masses $\sim 1/\ell$, carrying non-zero momentum
on $T^{2}$.       
\paragraph{}
To study the above theory in DLCQ, we will compactify the null
direction $x_{-}=x_{0}-x_{1}$ on a circle of radius $R_{-}$. The
conjugate momentum is therefore quantised: $p_{+}=K/R_{-}$ with 
$K\in \mathbb{N}$. According to the proposal of 
\cite{Ganor:Sethi:1997, Kapustin:Sethi:1998}, 
the sector of the theory 
with $K$ units of null momentum is described by maximally supersymmetric 
quantum mechanics on
the moduli space of $K$ Yang-Mills instantons of an auxiliary
$SU(N)$ gauge theory on $\mathbb{R}^{2}\times T^{2}$. This 
coincides with the quantum mechanical $\sigma$-model with target space
$\mathcal{M}$ described in the previous sections. 
The remaining non-compact light-cone coordinate
$x_{+}=x_{0}+x_{1}$ plays the role of time and the conjugate momentum
is identified with the $\sigma$-model Hamiltonian. Restoring
appropriate dimensionful factors we have, 
\begin{eqnarray}
p_{-} & = & H \nonumber \\  
& = & R_{-} {\rm Im}\tau_{\rm cl}\, \left[\inv^{IJ}\Pi_I
\bar{\Pi}_{J} \,+\, \frac{4}{\ell^{2}} \im \tau_{IJ}P^{I}
\bar{P}^{J}\,\,+\,\,{\rm fermions}\right].
\label{new2} 
\end{eqnarray}   
Here we have rescaled the base coordinates $a^{I}$ so that they have the
dimensions of length as appropriate for the DLCQ interpretation. 
More precisely, this $\sigma$-model Hamiltonian (\ref{new2}) 
describes physics in the region of the
moduli space where the metric can be replaced by its semi-flat form 
(\ref{eqn:Gsf}). Roughly speaking, this corresponds to studying the
theory at length-scales much larger than the compactification
scale $\ell$. This region should be the relevant one for understanding
the DLCQ description of ${\cal N}=4$ supersymmetric Yang-Mills theory.   
\paragraph{}
In addition to $\tau_{\rm cl}$, $R_{-}$ and $\ell$, 
the $\sigma$-model also depends on $N-1$ complex 
parameters corresponding to the relative positions of 
$N$ marked points $z=Z_{j}$ 
on a torus of complex structure $\tau_{\rm cl}$. In DLCQ,
these parameters correspond to a complex combination of light-like
electric and magnetic Wilson lines in the Cartan subalgebra of
$SU(N)$. Thus, when the $N$ points are distinct, the DLCQ description
is in a Coulomb phase where the low-energy gauge group is
$U(1)^{N-1}$. 
\paragraph{}
As above, the 
interactions of the model are determined by the period matrix 
$\tau_{IJ}=\partial^{2}\mathcal{F}/\partial a^{I}\partial a^{J}$ of
the complex curve $\Sigma$. 
The presence of Wilson lines provides a natural weak-coupling
expansion for the $\sigma$-model. It is convenient to choose the
light-like Wilson lines so that 
$\tau_{j}=(Z_{j+1}-Z_{j})/2\pi i= \alpha_{j}\tau_{\rm cl}$, for 
$j=1,2,\ldots,N$, where the variables $\{\alpha_{j}\}$ provide a
partition of unity,  
\begin{eqnarray}
0\leq \alpha_{j}\leq 1 & \qquad & \sum_{j=1}^{N}\alpha_{j}=1 .
\nonumber 
\end{eqnarray} 
To obtain a weak-coupling limit we
take ${\rm Im}\tau_{\rm cl}>>1$ with the parameters $\{\alpha_{j}\}$ held
fixed. In this limit we may replace the prepotential by its classical
value (\ref{cl}) and the target space $\mathcal{M}$ reduces
to,   
\begin{eqnarray}
\mathcal{M}_{\rm cl} & = & \mathbb{R}^{2}\times T^{2}[\tau_{\rm cl}] \times 
\prod_{j=1}^{N}\, \frac{ {\rm Sym}^{K}\left(\mathbb{R}^{2}\times 
T^{2}[\alpha_{j}\tau_{\rm cl}]\right)}{\mathbb{R}^{2}\times 
T^{2}[\alpha_{j}\tau_{\rm cl}]}  
\nonumber 
\end{eqnarray}
with a flat metric. In terms of the weak-coupling coordinates $\{a^{ib}\}$
introduced in the previous section we have, 
\begin{eqnarray}
p_{-} & = & H \nonumber \\  
& = & 
\sum_{j=1}^{N}\sum_{b=1}^{K}\, \frac{R_{-}}{\alpha_{j}}\, \left[  
\Pi^{jb}\bar{\Pi}^{jb}\,+\, \frac{1}{4\pi^{2}\ell^{2}} 
\left|Q_{e}^{jb}-\alpha_{j}\bar{\tau}_{\rm cl}Q_{m}^{jb}\right|^{2}\right]
\label{Hfree}
\end{eqnarray}  
Here $\Pi^{jb}=(\alpha_{j}/R_{-})\dot{\bar{a}}^{jb}$ is the
``covariant'' momentum conjugate to $a^{jb}$. $Q^{jb}_{e}$ and
$Q^{jb}_{m}$ are the conserved fibre momenta in the new coordinates
which are also subject to the constraints, 
\begin{eqnarray}
\frac{1}{N}\sum_{b=1}^{K}Q_{e}^{jb}\,=\,\bar{Q}_{e} & \qquad{} &    
\frac{1}{N}\sum_{b=1}^{K}Q_{m}^{jb}\,=\, \bar{Q}_{m}
\nonumber 
\end{eqnarray}
for $j=1,2,\ldots,N$. In the context of DLCQ, these constraints
imply the freezing out of the centre of mass for each species of parton, 
reflecting IR divergences coming from the logarithmic growth of the
Coulomb potential in two spatial dimensions \cite{Ganor:Sethi:1997}.  
The $2\pi$ periodicity of the fibre coordinates
$\{\theta^{I}_{e},\theta_{m,I}\}$ implies
that each component of $Q_{e}$ and $Q_{m}$ is quantised in integer
units\footnote{We set $\hbar=1$ throughout.}. 
\paragraph{}
To interpret this free limit of the model we start with the case
of vanishing fibre momentum; $Q^{jb}_{e}=Q^{jb}_{m}=0$. In this case
we recognise (\ref{Hfree}) as the Hamiltonian for $KN$
non-relativistic particles moving freely on $\mathbb{R}^{2}$. Each
Yang-Mills instanton has split up into $K$ constitutents which we
will call partons. There are $K$
identical partons in each of $N$ species. Each parton of the
$j$'th species has mass $\alpha_{j}/R_{-}$. To compare with the ${\cal
  N}=4$ theory, we recall that, in DLCQ, 
a free relativistic particle behaves like a non-relativistic
particle of mass $p_{+}$ 
moving in two non-compact dimensions transverse to the light-cone. 
In the presence of the light-like Wilson
lines described above, the quantisation rule for the null momentum 
$p_{+}$ carried by each particle is modified for particles charged
under the Cartan subgroup of $SU(N)$. For a particle carrying unit
charge under the subgroup $U(1)_{j}$ corresponding to the simple root
\footnote{Here $e_{j}$ denotes a matrix whose only non-zero entry is
  unity in the $j$'th position on the diagonal and we work with the
convention $e_{N+1}=e_{1}$.} $e_{j+1}-e_{j}$ of $SU(N)$, 
the quantisation rule is,  
\begin{eqnarray}
p_{+} & = & \frac{\alpha_{j}+K_{j}}{R_{-}} \qquad{} K_{j}\in\mathbb{N}.
\nonumber 
\end{eqnarray}
Thus it is consistent for a single unit of $p_{+}$ to be shared between 
$N$ particles where the $j$'th particle carries unit charge under 
$U(1)_{j}$. Putting these observations
together, we see that (\ref{Hfree}) describes the dynamics associated
with $K$ units of $p_{+}$, each shared between $N$ partons in this
way. In other words, the model correctly reproduces the DLCQ of the
free ${\cal N}=4$ theory. 
\paragraph{}
To complete the identification with the free ${\cal N}=4$ theory we
note that, in the full $\sigma$-model, the bosonic coordinates $a$,
$z$ of each parton are accompanied by fermions $\psi^{A}$ and 
$\bar{\psi}^{\bar{A}}$ in the ${\bf 4}$ and $\bar{\bf 4}$ of the
enhanced $SU(4)$ R-symmetry of the zero momentum sector. The fermions
have canonical anti-commutation relations 
$\{\psi^{A},\bar{\psi}^{\bar{A}}\}\sim \delta^{A\bar{B}}$ and give rise
to a fermionic Fock space in the usual way. Choosing a vacuum state 
$|0\rangle$
annihilated by all the $\bar{\psi}^{\bar{A}}$, we find a tower of
states associated with each parton, 
\begin{eqnarray} 
|0\rangle, \qquad{} \psi^{A}|0\rangle, 
\qquad{}\psi^{A}\psi^{B}|0\rangle, \qquad{}
\psi^{A}\psi^{B}\psi^{C}|0\rangle, \qquad{}
\psi^{A}\psi^{B}\psi^{C}\psi^{D}|0\rangle && \nonumber 
\end{eqnarray}
in the ${\bf 1}\oplus{\bf 4}\oplus{\bf 6}\oplus\bar{\bf 4}\oplus{\bf
  1}$ of $SU(4)$. These states fill out the usual on-shell light-cone 
supermultiplet of the ${\cal N}=4$ theory consisting of six scalars
of helicity zero, four fermion components each of helicity $\pm 1/2$
and the two transverse polarisations of the photon with helicity $\pm
1$. With a suitable charge assignment for the vacuum $|0\rangle$, the 
helicity corresponds to the $U(1)$ generator $\mathcal{R}$ of the
${\bf su}(1,1|4)$ light-cone superconformal algebra given in
(\ref{Rsym}) above.    
\paragraph{} 
Modes with non-zero values of the fibre momenta 
$Q_{e}$ and/or $Q_{m}$ have light-cone energy $p_{-}$ which 
scales like $1/\ell^{2}$. They are naturally associated with the
Kaluza-Klein modes of the $(2,0)$-theory carrying momenta on the
compactification torus. From the point of view of the low-energy gauge
theory, modes with $Q_{e}\neq 0$ are electrically charged and remain
light at weak coupling. In contrast, modes with $Q_{m}\neq 0$ are
magnetically charged. Their masses are non-perturbative in $g^{2}$ and
they are very massive at weak coupling.  
In fact, each parton of the ${\cal N}=4$ 
theory has two infinite towers of Kaluza-Klein excitations which
suggests a description in terms of free particles moving on the
transverse $\mathbb{R}^{2}\times T^{2}$ of the full $(2,0)$ 
compactification. More precisely, if we focus on electrically charged
states only, we can bring the Hamiltonian to the form,   
\begin{eqnarray}
p_{-} & = & H \nonumber \\  
& = & 
\sum_{j=1}^{N}\sum_{b=1}^{K}\, \frac{R_{-}}{\alpha_{j}}\, \left[  
\Pi^{jb}\bar{\Pi}^{jb}\,+\,  
\left(P_{e}^{jb}\right)^{2}\right]
\label{Hfreeb}
\end{eqnarray} 
corresponding to $K$ free partons of mass $\alpha_{i}/R_{-}$, for each
$i=1,2,\ldots,N$ moving on $\mathbb{R}^{2}\times
S^{1}$ where the rescaled $S^{1}$ has radius $\ell$ and 
$P_{e}^{jb}=Q^{jb}_{e}/2\pi \ell$ is the corresponding conserved
momentum. This is consistent with the DLCQ description of five-dimensional
free $SU(N)$ gauge theory compactified on $S^{1}$. 
\paragraph{}
Now we turn our attention to the interacting theory decribed by the
full Hamiltonian (\ref{new2}). One general feature is the presence
of $2(KN-N+1)$ conserved charges corresponding to the components of
$Q_{e}$ and $Q_{m}$. These correspond to the Kaluza-Klein momenta
carried by the {\em individual} partons, which are obviously conserved
in a limit where the theory becomes free. The surprising feature here is
that they remain conserved in the interacting theory. Of course this
is true only to the extent that the physics is correctly described by 
the semi-flat metric. As discussed above, interactions coming from
one-loop and 4d  instanton contributions to the prepotential lead to
singularities in the semi-flat metric. To describe states with
wavefunctions supported near these points, we need to resolve the
singularities by including 3d instanton effects as in 
\cite{Gaiotto:Moore:Neitzke:2008}. These however violate the isometries
of the semi-flat metric and lead to non-conservation of 
the charges $Q_{e}$ and $Q_{m}$.   
\paragraph{}
At this point the question arises: what can we hope to calculate in
this model? The spectrum of the light-cone Hamiltonian $p_{-}$ is
continuous and is hard to interpret, especially in an interacting
conformal field theory. The dilatation operator, which is defined in
the sector of zero fibre momentum, should have a discrete spectrum but 
the corresponding wavefunctions are centred near the origin which is
highly singular. The results of the previous section suggest an
answer: the model can be effectively regulated by introducing a
worldline magnetic field proportional to the holomorphic symplectic
form $\eta$ on ${\cal M}$. 
This corresponds to changing the light-cone Hamilonian $H$ in
(\ref{new2}) by the replacement $P^{I} \rightarrow P^{I}-\kappa
a^{I}$ for some parameter $\kappa$ with the dimensions of mass. 
In the next section we 
will suggest a spacetime interpretation for this modification of the 
DLCQ Hamiltonian. We define, 
\begin{eqnarray}
\mathbb{H} & = & \left.\frac{1}{\mu}
H\right|_{P^{I} \rightarrow P^{I}\,-\,\kappa a^{I}}
\nonumber 
\end{eqnarray}
where $\kappa$ and $\mu$ are two arbitrary mass scales. 
If we choose 
$\kappa=\ell \mu/2R_{-}$ then, for $P^{I}=0$, $\mathbb{H}$ reduces
to, 
\begin{eqnarray}
\mathcal{L}_{0}[\mu] & = &  \frac{1}{\mu}H+\mu K \nonumber \\ 
& = & 
\frac{R_{-}}{\mu} \, \inv^{IJ}\Pi_I
\bar{\Pi}_{J} \,+\, \frac{\mu}{R_{-}} 
\im \tau_{IJ}a^{I}\bar{a}^{J}\,\,+\,\,{\rm fermions}
\nonumber 
\end{eqnarray}  
which is isospectral to the dilatation operator of $SU(1,1|4)$ for any
choice of the scale $\mu$. The eigenstates of this 
operator are localised near the origin 
and, for these states, the problem of singularities remains. 
\paragraph{}
After a further rescaling $a^{I}\rightarrow a^{I}/\kappa$, to pass to
dimensionless coordinates, the new
Hamiltonian is given in full as\footnote{The shifting of the
  fermionic terms in the action follows from the homogeneity relation
  (\ref{eqn:conformalSK}) for the prepotential ${\cal F}(a)$.}, 
 \begin{align} \nonumber \begin{aligned} \mathbb{H} &= \frac{1}{\rho}\left[
\inv^{I\bar{J}}\Pi_I
\bar{\Pi}_{\bar{J}}
+ 
\frac{1}{2}R_{I\bar{J}K\bar{L}}\psi^{IA}\bar{\psi}^{\bar{J}}_A\psi^{KB}
\bar{\psi}^{\bar{L}}_B \right.
\\ & \left. + \frac{1}{12}\re 
\left(\epsilon_{ABCD}G_{IJKL}\psi^{IA}\psi^{JB}\psi^{KC}\psi^{LD}\right) \right]
\\ &+ \rho\,\im
\tau_{I\bar{J}}\left(P^{I}-a^{I}\right)\left(\bar{P}^{J}-
\bar{a}^{J}\right)
+ \frac{1}{2}\re 
\left( \F^{(3)}_{IJK}\Omega_{AB}\psi^{JA}\psi^{KB}\left(P^{I}-a^{I}\right)\right)
\end{aligned} \end{align}
where $\rho=4R_{-}/\mu\ell^{2}$. The above relations tell us that  
states with generic non-zero values of the fibre momenta
$P^{I}$ are exponentially localised near the (generically) non-singular point 
$a^{I}=P^{I}$. Moreover, the localisation of the wavefunction
is controlled by the dimensionless parameter $\rho$. It is
particularly interesting to consider the limit $\rho\rightarrow
\infty$, where the localisation is absolute and the relation
$a^{I}=P^{I}$ is imposed as a constraint on the phase space for all
states of finite energy. As a consequence of the constraint, the
coordinates 
$a^{I}$ and $z_{I}$ effectively become canonically conjugate
variables. The limit therefore reduces the full phase space of
the $\sigma$-model to the phase space of the complex integrable system 
$(\cal{M},J,\eta)$ described in Section 2. In leading order
semiclassical quantisation we consider wavefunctions of the form, 
\begin{eqnarray}
\Psi(z,\bar{z}) & = & \exp\left(i a^{I}z_{I}\right)
\exp\left(i \bar{a}^{I}\bar{z}_{I}\right).    
\nonumber 
\end{eqnarray}
The Bohr-Sommerfeld condition corresponds to demanding that these
wavefunctions are single-valued on the fibre ${\cal J}(\Sigma)$. Thus
we find quantisation conditions,  
\begin{eqnarray}
a^{I}+\bar{a}^{I} \in \mathbb{Z} & \qquad{} &
\tau_{IJ}a^{J}+\bar{\tau}_{IJ}\bar{a}^{J}  \in \mathbb{Z}
\nonumber 
\end{eqnarray}
or, more simply\footnote{ Note that for a scale-invariant
  prepotential, the homogeneity relation (\ref{eqn:conformalSK})
  implies $a^{D}_{I}=\partial{\cal F}/\partial a^{I}=\tau_{IJ}a^{J}$.} 
\begin{eqnarray}
2{\rm Re}\left[a^{I}\right] \in \mathbb{Z} & \qquad{} &
2{\rm Re}\left[a^{D}_{I}\right] \in \mathbb{Z}
\nonumber 
\end{eqnarray}
for $I=1,2,\ldots,r=KN-N+1$. The resulting spectrum is discrete and is
also invariant under the $Sp(2r,\mathbb{Z})$ monodromy group of the
periods.   
\paragraph{}
In the general case, the semiclassical quantisation conditions given
above look unfamiliar. In particular, they differ from any of the
quantisations of the same classical integrable system discussed in
\cite{Nekrasov:Shatashvili:2009} which involve the quantisation of {\em either} the
A-periods of the meromorphic differential $\lambda$ {\em or} the
B-periods (but not both). However, at least in the free limit of the
${\cal N}=4$ theory $\tau\rightarrow i\infty$ (with arbitrary fixed light-like
Wilson lines), we can make contact with a known quantum 
integrable system. In this limit,
the classical system becomes an $SL(N,\mathbb{C})$ Heisenberg spin
chain. The semiclassical quantisation conditions described above
coincide exactly with the semiclassical limit of a {\em quantum}
integrable $SL(N,{\mathbb C})$ spin chain with a principal series
representation at each site. In particular, the $N=2$ case is worked
out in detail in \cite{Derkachov:Korchemsky:Manashov:2002}.       
\paragraph{}
In the Lagrangian formulation, the same limit leads to a first-order
Lagrangian of the form, 
\begin{eqnarray}
{\cal L} & = & \frac{1}{2}\eta_{uv}w^{u}\dot{w}^{v}\,+\,
\frac{1}{2}\bar{\eta}_{\bar{u}\bar{v}}\bar{w}^{\bar{u}}\dot{\bar{w}}^{\bar{v}}
\nonumber \\ 
& = & a^{I}\dot{z}_{I}\,+\, \bar{a}^{I}\dot{\bar{z}}_{I}
\nonumber 
\end{eqnarray}
where the first line gives the Lagrangian in manifestly covariant form
where $w_{u}$, $\bar{w}_{\bar{u}}$, with $u,
\bar{u}=1,2,\ldots,2r$, are generic holomorphic coordinates 
on ${\cal M}$ for complex structure $J$. The second equality holds up
to surface terms.  
\paragraph{}
Finally we note that the effect of the
$\rho\rightarrow \infty$ limit is similar to that of the familiar limit which
isolates the lowest Landau level for a particle in a background
magnetic field \cite{Landau:Lifshitz}. This can be made explicit in the
weak-coupling limit discussed above. Before the introducing the deformation, the
electrically charged DLCQ partons (with $Q_{m}=0$)  
are governed by the Hamiltonian
(\ref{Hfreeb}) for free motion on $\mathbb{R}^{2}\times S^{1}$. 
The new Hamiltonian has the form,    
\begin{eqnarray}
p_{-} & = & H \nonumber \\  
& = & 
\sum_{j=1}^{N}\sum_{b=1}^{K}\, \frac{R_{-}}{\alpha_{j}}\, \left[  
\Pi^{jb}\bar{\Pi}^{jb}\,+\,  
\left|P_{e}^{jb}-\alpha_{j}\tilde{\kappa}a^{jb}\right|^{2}\right]
\label{Hfree2}
\end{eqnarray} 
where, as above, $P_{e}^{jb}=Q^{jb}_{e}/2\pi \ell$ is the conserved 
momentum around the electric cycle of $T^{2}$ and $\tilde{\kappa}$ is
a constant. Thus we see that the partons are moving in the presence of 
a background gauge field, where the light-cone momentum fraction $\alpha_{i}$
plays the role of electric charge.

\section{Spacetime interpretation}
\paragraph{}
In the preceding sections we have chosen to modify the $\sigma$-model
arising in the DLCQ description of 
${\cal N}=4$ supersymmetric Yang-Mills theory by introducing a
target space magnetic field. In this section we will propose a spacetime
interpretation of this modification using the ideas of
\cite{Maldacena:Martelli:Tachikawa:2008,Duval:Horvathy:Palla:1994}.   
\paragraph{}
We start from the ${\cal N}=4$ theory defined on Minkowski space 
$\mathbb{R}^{1,3}$ with light-cone coordinates $x_{\pm}=x_{0}\pm
x_{1}$ and transverse directions parametrised by a complex coordinate 
$u=x_{2}+ix_{3}$. The
spacetime metric is, 
\begin{eqnarray}
ds^{2}=dx_{+}dx_{-}+du d\bar{u}. 
\nonumber
\end{eqnarray}
As above we
choose a longitudinal null direction $x_{-}$ which is compactified on
a circle of radius $R_{-}$ in DLCQ. The coordinate $x_{+}$ is
interpreted as `time' and the light-cone Hamiltonian corresponds to
the conjugate momentum $p_{-}$. A free massless particle in Minkowski
space obeys the mass-shell condition $p_{+}p_{-}-\Pi_{u}\bar{\Pi}_{u}=0$,
where $\Pi_{u}$ is the conjugate momentum to $u$, which gives, 
\begin{eqnarray}
H=p_{-} & = & \frac{\Pi_{u}\bar{\Pi}_{u}}{p_{+}}. 
\nonumber 
\end{eqnarray}
This is the Hamiltonian for a non-relativistic particle of mass 
$p_{+}$. In a conformal theory the Hamiltonian is part of an $SO(2,1)$ group of
conformal transformations which also includes a light-cone dilatation
operator $T=D+M_{01}$ and special conformal transformation 
$K=K_{0}+K_{1}$ with commutation relations, 
\begin{eqnarray}
[T,K]\,=\,2iK \qquad{}  & [T,H]\,=\,-2iH  & \qquad{} [H,K]=-4iT.
\nonumber 
\end{eqnarray}
For a free massless particle the additional generators are
$T=u\Pi_{u}+\bar{u}\bar{\Pi}_{u}$ and $K=p_{+}u \bar{u}$.  
\paragraph{}
Minkowski space is conformally equivalent to the four-dimensional
pp-wave geometry with metric, 
\begin{eqnarray}
ds^{2}=dx_{+}dx_{-}+du d\bar{u} -\mu^{2}|u|^{2}dx_{+}^{2}
\label{4dpp}
\end{eqnarray}
where $\mu$ is an arbitrary mass scale. The pp-wave geometry has 
null isometries corresponding to shifts of $x_{+}$ and $x_{-}$. As
before we take $x_{+}$ as light-cone time and the conjugate momentum $p_{-}$
as the Hamiltonian. 
\paragraph{}
The conformal map which relates the pp-wave to
Minkowski space maps the conformal generators of one space onto the
other. According to \cite{Maldacena:Martelli:Tachikawa:2008,Duval:Horvathy:Palla:1994} we have, 
\begin{eqnarray}
H_{\rm pp-wave} & = & H\,+\, \mu^{2}K.
\label{eqnarray}
\end{eqnarray}
Up to an overall factor of $\mu$, it follows that $H_{\rm pp-wave}$
has the same spectrum as the Minkowski space light-cone dilatation operator
$T$. The conformal equivalence means that, up to global issues, the
${\cal N}=4$ theory on the pp-wave is equivalent to the theory on
Minkowski space with the appropriate identification of conformal
generators. In particular, to determine the spectrum of 
the light-cone dilatation operator of the
${\cal N}=4$ theory in Minkowski space, one can equivalently study the light-cone
Hamiltonian of the ${\cal N}=4$ theory in the plane wave geometry
(\ref{4dpp})\footnote{This statement can be compared with the more familiar
equivalence of the dilatation operator on $\mathbb{R}^{1,3}$ and the
Hamiltonian of the same theory on $\mathbb{R}\times S^{3}$ provided
by radial quantisation.}.   
\paragraph{}
To construct the ${\cal N}=4$ theory on a pp-wave geometry one can 
hope to start from the $(2,0)$ theory on a six-dimensional
space which reduces at large distances to a four-dimensional pp-wave after
compactification on a torus.  To realise this idea, 
we start by considering six-dimensional Minkowski space in light-cone
coordinates $x_{\pm}=x_{0}\pm x_{1}$ where the transverse space is
parametrised by complex coordinates $u=x_{2}+ix_{3}$ and
$v=x_{4}-ix_{5}$. The Minkowski space metric is 
\begin{eqnarray}
ds^{2} & = & dx_{+}dx_{-}+du d\bar{u}+ dvd\bar{v}. 
\nonumber
\end{eqnarray}
To obtain a four-dimensional theory we compactify $x_{4}$ and $x_{5}$
on circles of radius $R_{4}$ and $R_{5}$ respectively, giving
the identifications, 
\begin{eqnarray}
v \sim v + 2\pi R_{4} & \qquad{} & v \sim v + 2\pi i R_{5}
\nonumber 
\end{eqnarray}
To obtain $\mathcal{N}=4$ SUSY Yang-Mills at low energies with
coupling $g^{2}$ (and $\theta=0$) we set 
$R_{5}/R_{4}={\rm Im}\tau_{\rm cl}=4\pi/g^{2}$. 
In light-cone coordinates the light-cone Hamiltonian for a free
particle in six-dimensions takes the form, 
\begin{eqnarray}
H=p_{-} & = & \frac{1}{p_{+}}\left[ \Pi_{u}\bar{\Pi}_{u}\,+\,  
\Pi_{v}\bar{\Pi}_{v}\right]
\nonumber
\end{eqnarray}
where $\Pi_{u}$ and $\Pi_{v}$ are conjugate momenta to the complex
coordinates $u$ and $v$ respectively. Thus, the free particle in six
dimensions becomes a free non-relativistic particle of mass $p_{+}$
moving on the transverse space $\mathbb{R}^{2}\times T^{2}$. This is
the starting point for a DLCQ description of a six-dimensional QFT 
compactified to four dimensions on $T^{2}$.    
\paragraph{}
In order to make contact with the ideas of the previous sections we
consider instead a six-dimensional space where the light-like
direction $x_{-}$ is fibred non-trivially over the other dimensions
with metric, 
 \begin{eqnarray}
ds^{2} & = & dx_{+}\left(dx_{-}+\mu udv+ \mu \bar{u}d\bar{v}\right) 
+du d\bar{u}+ dvd\bar{v} 
\label{6dm}
\end{eqnarray}
where $\mu$ is a mass scale. 
This is a Bargmann space of the type considered in \cite{Duval:Horvathy:Palla:1994}. The
resulting light-cone Hamiltonian takes the form, 
\begin{eqnarray}
\tilde{H}=p_{-} & = & \frac{1}{p_{+}}\Pi_{u}\bar{\Pi}_{u}\,+\,  
\frac{1}{p_{+}}
\left(\Pi_{v}-p_{+}\mu u\right)\left(\bar{\Pi}_{v}-p_{+}\mu \bar{u}\right).
\label{Barg}
\end{eqnarray}
Thus the effect of the fibration on a non-relativistic particle is to
give it an electric charge equal to its
mass $p_{+}$ and couple it to a background anti-self-dual magnetic field
, 
\begin{eqnarray}
f & = & \mu\left(du\wedge dv \, + \, d\bar{u}\wedge  d\bar{v}\right) .
\label{eqn:fieldstrength}
\end{eqnarray}
This matches the weak coupling limit (\ref{Hfree2}) of the DLCQ
Hamiltonian. Thus, the deformation
of the $\sigma$-model obtained by introducing target space magnetic
field is a natural candidate for the DLCQ description of the $(2,0)$
theory compactified on the Bargmann space (\ref{6dm}).
\paragraph{}
In the case that the fibre momentum vanishes, the Hamiltonian reduces
to, 
\begin{eqnarray}
H_{\rm pp-wave} & = & \frac{1}{p_{+}}\Pi_{u}\bar{\Pi}_{u}\,+\,  
\mu^{2} p_{+} u\bar{u}.
\nonumber
\end{eqnarray}
This can also be understood by completing the square to put the
six-dimensional metric (\ref{6dm}) in the form 
 \begin{eqnarray}
ds^{2} & = & dx_{+}dx_{-}+ 
du d\bar{u} -\mu^{2}|u|^{2}dx_{+}^{2} + \left(dv-\bar{u}dx_{+}\right)  
\left(d\bar{v}-u dx_{+}\right) 
\label{6dm2}
\end{eqnarray}
which manifestly has the form of a $T^{2}$ bundle over the
four-dimensional pp-wave geometry (\ref{4dpp}). 
\paragraph{}
We will support this interpretation further by analysing the effect of the
anti-self-dual magnetic
field (\ref{eqn:fieldstrength}), whose potential is
\[ a = \mu \left(udv+\bar{u}d\bar{v}\right),\]
in the DLCQ description of the $(2,0)$ theory compactified on 
$\mathbb{R}^{2}\times T^{2}$. Thus we investigate the
effect of this background field on slowly-moving Yang-Mills
instantons. Since the effective electric charge $p_+$ in (\ref{Barg}) is set by
the instanton number $p_+ = K/R_-$, the corresponding 
gauge potential $a$ must couple minimally to
the instanton current,
\[ j = \frac{1}{8\pi^2}*\tr F \wedge F.\]
The resulting five-dimensional action is the $\mathcal{N}=2$ supersymmetric
completion of
\begin{equation} S = -\frac{1}{g^2}\int_{\mathbb{R}^{2,1}\times T^2} \tr \left(F\wedge *F
  + F\wedge F\wedge a\right),\label{eqn:SwithA} \end{equation}
where $g^2 = R_-/8\pi^2$ and the normalisation is chosen so that a $K$-instanton
has electric charge $p_+ = K/R_-$. From the stringy
perspective this can be understood as a modification of the D4-brane action
due to a background Ramond-Ramond 1-form flux proportional to $a$.
\paragraph{}
A surprising fact is that, by virtue of the anti-self-duality of $f$,
a static, self-dual instanton is still an exact solution of the modified
equations of motion
\[ D_{\nu}F^{\nu \mu} - \frac{1}{4}\epsilon^{\mu \nu \rho
\sigma \tau}F_{\nu \rho}f_{\sigma \tau} = 0.\]
It therefore makes sense to work with the moduli space approximation. The
standard procedure is to promote the collective coordinates $X^{\alpha}$,
$\alpha = 1,\dots,4(KN-N+1)$,
to slowly varying dynamical degrees of freedom and set
\[ A_0 = \dot{X}^{\alpha}\Omega_{\alpha}, \qquad A_m = A_m^{inst}\left(x
;X^{\alpha}(t)\right) \quad m=1,2,3,4\]
where $\Omega_{\alpha}$ is the gauge-fixing parameter defined by
\[ D_m \delta_{\alpha} A_m := D_m\left(\partial_{\alpha}A_m
 - D_m \Omega_{\alpha}\right) = 0.\]
One finds that the equations of motion are solved to lowest order in time
derivatives. Substituting this ansatz into the action (\ref{eqn:SwithA})
leads to
\[ S = \frac{1}{g^2}\int dt -\frac{1}{2}\mathcal{G}_{\alpha \beta}\dot{X}^{\alpha}
\dot{X}^{\beta} + \mathcal{A}_{\alpha}\dot{X}^{\alpha},\]
where
\[ \mathcal{G}_{\alpha \beta} = -2\int d^4x\tr
 \left(\delta_{\alpha}A_m\delta_{\beta}A_m\right)\]
is the usual metric on moduli space and we interpret
\[ \mathcal{A}_{\alpha} = -2\int d^4x
 \tr \left(\delta_{\alpha}A_m F_{mn} a_n\right)\]
as the instanton worldline magnetic field induced by $a$.
\paragraph{}
The field strength associated to $\mathcal{A}$ is
\[ \mathcal{F}_{\alpha \beta} = -2\int d^4
 x \tr \left(\delta_{\alpha}A_m f_{mn}\delta_{\beta}A_n\right).\]
But the holomorphic symplectic form on $\mathbb{R}^2 \times T^2$ (with respect
to the preferred complex structure induced by holomorphic coordinates
$u$ and $v$) is just $du \wedge dv$. By the usual rules for inducing the
hyper-K\"{a}hler structure on moduli space from that of Euclidean space,
the corresponding holomorphic symplectic form $\eta$ on moduli space is
given in terms of $f$ and $\mathcal{F}$ by
\[ \mathcal{F}_{\alpha \beta} = \mu \left(\eta + \bar{\eta}\right)_
{\alpha \beta} = -2\int d^4x
\tr\left(\delta_{\alpha}A_m f_{mn} \delta_{\beta}A_n\right). \]
We conclude that the
effect of the spacetime magnetic field arising from Kaluza-Klein reduction
of the Bargmann geometry (\ref{6dm}), is to produce exactly
the deformation used in section 3 to regularise the DLCQ description of
the $\mathcal{N} = 4$ theory.

\section{Discussion}
\paragraph{}
In this paper we have presented further evidence in favour of the DLCQ
description of the ${\cal N}=4$ theory proposed in 
\cite{Ganor:Sethi:1997, Kapustin:Sethi:1998}. The most important
question however is whether and to what extent 
the model can be used to calculate gauge theory observables. An
obvious goal is to reproduce the planar anomalous dimensions of the
${\cal N}=4$ theory, hopefully shedding new light on the underlying
integrable structure. Although our results are preliminary, they
suggest a possible approach to this problem. The idea is to work with
the deformed model in the $\rho\rightarrow \infty$ limit where we have 
already established semiclassical integrability. A full treatment
would also include the fluctuations of both bosonic and fermionic
fields around the leading order solution discussed above. Next one
seeks a quantisation of this system which preserves the semiclassical
integrability. The weak coupling limit described above where the
system can be related to a known spin chain looks particularly
promising in this regard. Interestingly, in addition to a na\"{i}ve weak
coupling limit $\tau\rightarrow i\infty$, the connection to a spin
chain seems to hold also in a 't Hooft limit where $N$ is also taken
to infinity holding $N/{\rm Im}\tau$ fixed. 
\paragraph{}
The next step is to focus on states in the sector with $Q_{e}=Q_{m}=0$
corresponding to the gauge-invariant 
states of the ${\cal N}=4$ theory. In fact 
the integrable system arising {\em after} taking the
$\rho\rightarrow\infty$ limit seems to extend smoothly to this sector
despite the fact that the model is singular for states with 
$Q_{e}=Q_{m}=0$ for any finite value of $\rho$.   
Of course, if this
approach is correct, we would expect to find some obstruction to
integrability away from the large-$N$ 't Hooft limit. 
We would also expect to find some
relation between the quantum corrected eigenvalues of the 
dilatation operator and the conserved charges of the integrable
system. The final step is to take a $K\rightarrow \infty$ limit to
decompactify the light-like circle and 
compare with the ${\cal N}=4$ theory in Minkowski space.   
These issues are currently under investigation \cite{ND:future}.        
\paragraph{}
We would like to thank Sungjay Lee and David Tong for useful
discussions. The research leading to these results has received 
funding from the European Research Council under the European
Community's Seventh Framework Programme (FP7/2007-2013) / 
ERC grant agreement no. [247252]. AS is supported by the STFC.

\bibliography{IIDLCQ}

\providecommand{\href}[2]{#2}\begingroup\raggedright\begin{thebibliography}{10}

\bibitem{BigReview:2010}
N.~Beisert, C.~Ahn, L.~Alday, Z.~Bajnok, J.~Drummond, et~al., {\it {Review of
  AdS/CFT Integrability: An Overview}},  {\em Lett.Math.Phys.} {\bf 99} (2012)
  3--32, [\href{http://xxx.lanl.gov/abs/1012.3982}{{\tt arXiv:1012.3982}}].

\bibitem{Delduc:Magro:Vicedo:2012}
F.~Delduc, M.~Magro, and B.~Vicedo, {\it {Alleviating the non-ultralocality of
  the $AdS_5 \times S^5$ superstring}},  {\em JHEP} {\bf 1210} (2012) 061,
  [\href{http://xxx.lanl.gov/abs/1206.6050}{{\tt arXiv:1206.6050}}].

\bibitem{Ganor:Sethi:1997}
O.~Ganor and S.~Sethi, {\it {New perspectives on Yang-Mills theories with
  sixteen supersymmetries}},  {\em JHEP} {\bf 9801} (1998) 007,
  [\href{http://xxx.lanl.gov/abs/hep-th/9712071}{{\tt hep-th/9712071}}].

\bibitem{Kapustin:Sethi:1998}
A.~Kapustin and S.~Sethi, {\it {The Higgs branch of impurity theories}},  {\em
  Adv.Theor.Math.Phys.} {\bf 2} (1998) 571--591,
  [\href{http://xxx.lanl.gov/abs/hep-th/9804027}{{\tt hep-th/9804027}}].

\bibitem{Witten:1995}
E.~Witten, {\it {Some comments on string dynamics}},
  \href{http://xxx.lanl.gov/abs/hep-th/9507121}{{\tt hep-th/9507121}}.

\bibitem{ND:AS:2014}
N.~Dorey and A.~Singleton, {\it {Superconformal Quantum Mechanics and the
  Discrete Light-Cone Quantisation of N=4 SUSY Yang-Mills}},
  \href{http://xxx.lanl.gov/abs/1409.8440}{{\tt arXiv:1409.8440}}.

\bibitem{ND:future}
N.~Dorey, {\it {Work in progress}}, .

\bibitem{Nekrasov:1995}
N.~Nekrasov, {\it {Holomorphic bundles and many body systems}},  {\em
  Commun.Math.Phys.} {\bf 180} (1996) 587--604,
  [\href{http://xxx.lanl.gov/abs/hep-th/9503157}{{\tt hep-th/9503157}}].

\bibitem{Seiberg:Witten:1996}
N.~Seiberg and E.~Witten, {\it {Gauge dynamics and compactification to
  three-dimensions}},  \href{http://xxx.lanl.gov/abs/hep-th/9607163}{{\tt
  hep-th/9607163}}.

\bibitem{Kapustin:1998}
A.~Kapustin, {\it {Solution of N=2 gauge theories via compactification to
  three-dimensions}},  {\em Nucl.Phys.} {\bf B534} (1998) 531--545,
  [\href{http://xxx.lanl.gov/abs/hep-th/9804069}{{\tt hep-th/9804069}}].

\bibitem{ND:Hollowood:Kumar:2001}
N.~Dorey, T.~Hollowood, and S.~Kumar, {\it {An Exact elliptic superpotential
  for N=1* deformations of finite N=2 gauge theories}},  {\em Nucl.Phys.} {\bf
  B624} (2002) 95--145, [\href{http://xxx.lanl.gov/abs/hep-th/0108221}{{\tt
  hep-th/0108221}}].

\bibitem{Intriligator:Seiberg:1996}
K.~Intriligator and N.~Seiberg, {\it {Mirror symmetry in three-dimensional
  gauge theories}},  {\em Phys.Lett.} {\bf B387} (1996) 513--519,
  [\href{http://xxx.lanl.gov/abs/hep-th/9607207}{{\tt hep-th/9607207}}].

\bibitem{Witten:1997}
E.~Witten, {\it {Solutions of four-dimensional field theories via M theory}},
  {\em Nucl.Phys.} {\bf B500} (1997) 3--42,
  [\href{http://xxx.lanl.gov/abs/hep-th/9703166}{{\tt hep-th/9703166}}].

\bibitem{Gaiotto:Moore:Neitzke:2008}
D.~Gaiotto, G.~Moore, and A.~Neitzke, {\it {Four-dimensional wall-crossing via
  three-dimensional field theory}},  {\em Commun.Math.Phys.} {\bf 299} (2010)
  163--224, [\href{http://xxx.lanl.gov/abs/0807.4723}{{\tt arXiv:0807.4723}}].

\bibitem{Lee:Yi:1997}
K.-M. Lee and P.~Yi, {\it {Monopoles and instantons on partially compactified
  D-branes}},  {\em Phys.Rev.} {\bf D56} (1997) 3711--3717,
  [\href{http://xxx.lanl.gov/abs/hep-th/9702107}{{\tt hep-th/9702107}}].

\bibitem{Luty:Taylor:1995}
M.~Luty and W.~Taylor, {\it {Varieties of vacua in classical supersymmetric
  gauge theories}},  {\em Phys.Rev.} {\bf D53} (1996) 3399--3405,
  [\href{http://xxx.lanl.gov/abs/hep-th/9506098}{{\tt hep-th/9506098}}].

\bibitem{Kricheveretal:1994}
I.~Krichever, O.~Babelon, E.~Billey, and M.~Talon, {\it {Spin generalization of
  the Calogero-Moser system and the matrix KP equation}},
  \href{http://xxx.lanl.gov/abs/hep-th/9411160}{{\tt hep-th/9411160}}.

\bibitem{Gorskyetal:1996}
A.~Gorsky, A.~Marshakov, A.~Mironov, and A.~Morozov, {\it {N=2 supersymmetric
  QCD and integrable spin chains: Rational case $N(f) < 2N(c)$}},  {\em
  Phys.Lett.} {\bf B380} (1996) 75--80,
  [\href{http://xxx.lanl.gov/abs/hep-th/9603140}{{\tt hep-th/9603140}}].

\bibitem{Gorsky:Gukov:Mironov:1997}
A.~Gorsky, S.~Gukov, and A.~Mironov, {\it {Multiscale N=2 SUSY field theories,
  integrable systems and their stringy / brane origin. 1.}},  {\em Nucl.Phys.}
  {\bf B517} (1998) 409--461,
  [\href{http://xxx.lanl.gov/abs/hep-th/9707120}{{\tt hep-th/9707120}}].

\bibitem{AlvarezGaume:Freedman:1981}
L.~Alvarez-Gaum{\'e} and D.~Freedman, {\it {Geometrical structure and
  ultraviolet finiteness in the supersymmetric sigma model}},  {\em
  Commun.Math.Phys.} {\bf 80} (1981) 443.

\bibitem{FOF:Kohl:Spence:1997}
J.~Figueroa-O'Farrill, C.~K{\"o}hl, and B.~Spence, {\it {Supersymmetry and the
  cohomology of (hyper)K{\"a}hler manifolds}},  {\em Nucl.Phys. B} {\bf 503}
  (1997) 614--626, [\href{http://xxx.lanl.gov/abs/hep-th/9705161}{{\tt
  hep-th/9705161}}].

\bibitem{Verbitsky:1990}
M.~Verbitsky, {\it {Action of the Lie algebra SO(5) on the cohomology of a
  hyperk{\"a}hler manifold}},  {\em Functional Analysis and Its Applications}
  {\bf 24} (1990), no.~3 229--230.

\bibitem{Nishida:Son:2007}
Y.~Nishida and D.~Son, {\it {Nonrelativistic conformal field theories}},  {\em
  Phys.Rev.} {\bf D76} (2007) 086004,
  [\href{http://xxx.lanl.gov/abs/0706.3746}{{\tt arXiv:0706.3746}}].

\bibitem{Kim:Kim:Koh:Lee:Lee:2011}
H.-C. Kim, S.~Kim, E.~Koh, K.~Lee, and S.~Lee, {\it {On instantons as
  Kaluza-Klein modes of M5-branes}},  {\em JHEP} {\bf 1112} (2011) 031,
  [\href{http://xxx.lanl.gov/abs/1110.2175}{{\tt arXiv:1110.2175}}].

\bibitem{Nekrasov:Shatashvili:2009}
N.~Nekrasov and S.~Shatashvili, {\it {Quantization of Integrable Systems and
  Four Dimensional Gauge Theories}},
  \href{http://xxx.lanl.gov/abs/0908.4052}{{\tt arXiv:0908.4052}}.

\bibitem{Derkachov:Korchemsky:Manashov:2002}
S.~Derkachov, G.~Korchemsky, and A.~Manashov, {\it {Noncompact Heisenberg spin
  magnets from high-energy QCD. 3. Quasiclassical approach}},  {\em Nucl.Phys.}
  {\bf B661} (2003) 533--576,
  [\href{http://xxx.lanl.gov/abs/hep-th/0212169}{{\tt hep-th/0212169}}].

\bibitem{Landau:Lifshitz}
L.~Landau and E.~Lifshitz, {\em {Quantum Mechanics (Non-relativistic Theory)}}.
\newblock Butterworth-Heinmann, 1997.

\bibitem{Maldacena:Martelli:Tachikawa:2008}
J.~Maldacena, D.~Martelli, and Y.~Tachikawa, {\it {Comments on string theory
  backgrounds with non-relativistic conformal symmetry}},  {\em JHEP} {\bf
  0810} (2008) 072, [\href{http://xxx.lanl.gov/abs/0807.1100}{{\tt
  arXiv:0807.1100}}].

\bibitem{Duval:Horvathy:Palla:1994}
C.~Duval, P.~Horvathy, and L.~Palla, {\it {Conformal properties of Chern-Simons
  vortices in external fields}},  {\em Phys.Rev.} {\bf D50} (1994) 6658--6661,
  [\href{http://xxx.lanl.gov/abs/hep-th/9404047, hep-th/9405229}{{\tt
  hep-th/9404047, hep-th/9405229}}].

\end{thebibliography}\endgroup
\bibliographystyle{JHEP}

\end{document}